\documentclass[11pt,twoside]{article} 
\usepackage{times,fancyhdr}
\usepackage{cite}
\usepackage{graphicx}

\usepackage{amsmath}
\usepackage{amsfonts}
\usepackage{amssymb}

\usepackage{rotating}

\newtheorem{theorem}{Theorem}

\newtheorem{corollary}{Corollary}

\newtheorem{lemma}{Lemma}

\newtheorem{proposition}{Proposition}

\newenvironment{proof}[1][Proof]{\textbf{#1.} }{\ \rule{0.5em}{0.5em}}

\date{}

\title{Embeddings of Lie algebras, contractions and the state labelling problem} 
\author{R. Campoamor-Stursberg\\I.M.I and Universidad Complutense de Madrid,\\ Plaza de
Ciencias 3, E-28040 Madrid.\\ rutwig@pdi.ucm.es}

\begin{document} 

\fancyhead{} 
\fancyhead[EC]{author name} \fancyhead[EL,OR]{\thepage}
\fancyhead[OC]{article name}
\fancyfoot{} 
\renewcommand\headrulewidth{0.5pt}
\addtolength{\headheight}{2pt} 

\maketitle 

\section{Introduction}

Since its introduction in Quantum Mechanics, group theory has shown to be a powerful
tool to understand and interpret physical phenomena, from the crystalline
structure of solids and the interpretation of atomic spectra to the classification of
particles and the establishment of nuclei models. In all these applications,
the groups are related usually to the symmetries of the system, either as spectrum-generating
or dynamical groups, where the Casimir operators of the corresponding Lie algebra and those
of distinguished subalgebras play a central role to describe the Hamiltonian or construct
mass formulae. In this context, one of the main situations where group theoretical methods are applied to
physical problems is concerned with classification schemes, where
irreducible representations of a Lie  group have to be decomposed
into irreducible representations of a certain subgroup appearing
in some relevant reduction chain
\begin{equation}
\left|
\begin{array}
[c]{cccccccccc}%
 \frak{s} & \supset &  \frak{s}^{\prime} & \supset &  \frak{s}^{\prime\prime} & ...& \supset &  \frak{s}^{(n)} & ...\\
\downarrow &  & \downarrow &  & \downarrow & & & \downarrow & \\
 \left[  \lambda\right]   &  &  \left[
\lambda^{\prime}\right] & &  \left[ \lambda^{\prime\prime}\right]
& ...&\supset & \left[ \lambda^{(n)}\right] & ...
\end{array}
\right\rangle. \label{Red1}
\end{equation}
This is the case for dynamical symmetries used for example in
nuclear physics, where one objective of the algebraic model is to
describe the Hamiltonian (or mass operator in the relativistic
frame) as a function of the invariant operators of the chain
elements. The corresponding energy formulae can the easily deduced
from the expectation values in the reduced representations. As
example, the Gell-Mann-Okubo mass formula is derived using
this ansatz \cite{Ok}. In many situations, the labels obtained
from the reduction (\ref{Red1}) are sufficient to solve the
problem, e.g., if we require multiplicity free reductions, as used
in $SU(N)$ tumbling gauge models \cite{Sua} or the interacting
boson model \cite{Ia}. However, often the subgroup does not
provide a sufficient number of labels to specify the basis states
unambigously, and multiplicities greater than one appear in the induced
representations. This happens in many of the  non-canonical
embeddings and generic irreducible representations (IRREPs) of Lie algebras.
Often this is not a constraint, since the interesting representations
belong to a certain type, like totally symmetric or anti-symmetric
representations, and additional labels are not necessary to solve the
problem, the degeneracies being solved directly with the
available Casimir operators.

\medskip

Many different methods and procedures to solve the so-called
missing label problem (short MLP) have been developed in the literature,
like projection of states, construction of
states for the members of the reduction chain, the study of the enveloping
algebras to determine all possible labelling operators, etc \cite{El}.
Even if the latter procedure allows to find the most general
labelling operator, the effective computation of
integrity bases is a rather complicated problem, and no effective method
is available. Among the difficulties appearing in this
approach, we remark that no general criterion to decide how many
 operators are necessary to generate an integrity bases is known. From
 the pure physical point of view, the question whether the found operators
 have some intrinsic meaning remains open, the operators having been obtained
 by a formal procedure. It is however expectable that labelling operators must
 have some interpretation in a physical context, as happens for the Elliott chain
 $\frak{su}(3)\supset \frak{so}(3)$
used in atomic physics, the Racah chain $\frak{so}(7)\supset G_{2}\supset\frak{so}(3)$
used in the description of  $f$-electron configurations, the Interacting Boson
Model based on the spectrum generating unitary Lie algebra $\frak{u}(6)$ or
the K-matrix theory used in the nuclear $\frak{sp}(3)$ model
\cite{Ia,El,Ro,Ro2,Is,Fet}.

\medskip

Using the original conception of Lie groups as groups of transformations with their
infinitesimal generators, an analytical approach using differential equations is possible,
and easily adaptable to the MLP \cite{Pe}. This method generalizes previous
procedures to compute the Casimir invariants of Lie algebras, and corresponds to interpret
 Casimir operators as functions that are constant on co-adjoint
orbits. One of the advantages of the analytical ansatz is that it is nor restricted
to invariants of polynomial type. From this perspective, labelling operators can be seen as
particular solutions of a certain subsystem of partial differential equations corresponding to an
embedded subalgebra. Classical operators are recovered easily using the
symmetrization map for tensors. In principle, the analytical method is
more direct than the pure algebraic
approach based on enveloping algebra, although integration
of systems of differential equations is far from being
a trivial task. Additional constraints like the orthogonality of
labelling operators are still not expressible in analytical way.

\medskip

Generally, labelling problems have been analyzed for specific chains of algebras,
either combining some of the above mentioned techniques, or from an algebraic point
of view, determining the operators of lowest degree that solve the state labelling.
It is not unusual that a complete solution is still unknown, or that only
certain types of labelling operators have been constructed. This is usually justified
by the computational complexity of finding the general expression of the labelling
operators. Another point of view involves the use of the properties and branching rules
determined by each embedding of a Lie algebra into a larger one. \footnote{Since non-equivalent
 embeddings of algebras lead to different branching rules, i.e., different decompositions of
 induced representations, the corresponding classification schemes are completely different.}
In any application, the way how a subalgebra es embedded into a larger
symmetry algebra $\frak{s}\supset \frak{s}^{\prime}$ reflects the physics of
the problem, corresponding to a coupling scheme or some symmetry breaking. It is not unreasonable
to think that in the case of non-multiplicity free reductions, the labelling operators needed can
be deduced from the data provided by the embedding. Since symmetry breaking is known to be
related with contractions of Lie algebras \cite{Vi}, we can ask to which extent the labelling problem
 can be solved without using external formal machinery.
In this context, the preserved symmetry corresponds to
some subalgebra which remains unchanged by the contraction. We remark that this approach  underlies
the rotor expansion method developed in \cite{Ro2}.

\medskip

Assuming the relation of the missing label problem with contractions of Lie algebras, we can ask
under which conditions they provide the labelling operators with the required properties. Formulated in another
way: how many labelling operators of the reduction $\frak{s}\supset \frak{s}^{\prime}$ can be obtained
using the symmetry breaking with respect to $\frak{s}^{\prime}$?
The first version of this approach to the missing label problem was
developed in \cite{C72}, having in mind the characterization of inhomogeneous
algebras obtained
from contractions of semisimple algebras \cite{C49}. It was observed that
any reduction chain $\frak{s}\supset \frak{s}^{\prime}$ is naturally
related to some types of inhomogeneous Lie algebras obtained by a contraction procedure.
The next step was to give a certain meaning to
the invariants of the contraction, and their possible connection with solutions to
the corresponding MLP. This first development only considered the contracted invariants
to generate labelling operators. This approach sufficed to solve physically relevant missing label
problems, like those with one labelling operator, as well as other with a higher number. The results
 were in harmony with those obtained using other methods. The limits of validity of the method were
 also established, observing that for reductions satisfying the
identity $\mathcal{N}(\frak{g})=\mathcal{N}(\frak{s})=n$ no complete solution was available. This
failure is a consequence of an insufficient number of contracted invariants independent from the
initial Casimir operators. It was also remarked that, for some special cases, although a sufficient
number of independent solutions can found, linear combinations of these are not mutually orthogonal.

\medskip

The main objective of the generalized contraction ansatz in labelling
problems can be resumed in the following points:

\begin{enumerate}

\item Find an effective method to solve the  MLP using explicitly the
properties of the embedding  $\frak{s}\supset \frak{s}^{\prime}$ and the
decomposition it induces on the Casimir operators.

\item Justify a physical choice of labelling operators as
``broken Casimir operators" or linear functions of them.

\item Find a satisfactory explanation for the non-integer
expectation values of labelling operators observed in the classical reduction chains.

\end{enumerate}

\medskip

The aim of this work is to review the actual progress on the missing label problem using the
contraction ansatz, as well as some applications where this procedure could be of notable interest.
The extrapolation of this approach to other types of reduction chains, like the problem of the
Racah operators, is also outlined.  More
specifically, we combine the analytical method of \cite{Pe} for
solving the MLP with contractions of Lie algebras with some refinements
concerning the decomposition of Casimir operators. After recalling that
for any embedding $\frak{s}\supset\frak{s}^{\prime}$ of
(semisimple) Lie algebras we can find an associated simple
In\"on\"u-Wigner contraction of $\frak{s}$ onto an affine Lie
algebra
$\frak{g}=\frak{s}^{\prime}\overrightarrow{\oplus}_{R}nL_{1}$,
where $nL_{1}$ denotes an $n$-dimensional Abelian algebra and $R$
is a representation of the subalgebra $\frak{s}^{\prime}$ such
that the adjoint representation $ad$ of $\frak{s}$ satisfies the
condition $ad(\frak{s})=ad(\frak{s}^{\prime})\oplus R$, we see
that any invariant of the contraction $\frak{g}$
can be taken as a solution to the missing label operator. The first question
to be solved is whether and under which constraints the invariants of the contraction
$\frak{g}$ are sufficient in number to provide a set of missing
label operators. We give sufficient conditions  to solve the
MLP by means of this associated contraction, and derive some
useful consequences on their structure. At this point we observe that
 the missing label operators inherit a physical interpretation as
 the terms of Casimir operators that disappear during contraction. The
 possibility of recovering them by linear combinations show that
they are internally determined by the group-subgroup chain.
For some degenerate cases, where no missing labels exist, we observe that the
 invariants of the contraction arise as polynomial functions of the Casimir operators of the contracted
Lie algebra $\frak{s}$ and the subalgebra $\frak{s}^{\prime}$. Generalizing
this approach, for the
cases where the contraction alone is not sufficient to find a set of
labelling operators, a refinement based on a decomposition of the Casimir operators
is proposed. It is proven that any of the terms of this decomposition are solutions
to the labelling problem. This provides more possibilities to derive an
orthogonal set, and explains some features already observed in the literature, like
the minimal degree of the labelling operators. Finally, it is commented to which
extent this refinement holds, and possible future outlines are presented.

\section{Missing label operators}

It is well known from classical theory that any semisimple Lie algebra $\frak{g}$ possesses exactly
$\mathcal{N}(\frak{g})=l$ independent Casimir operators, i.e., polynomials in the generators
that commute with all elements of the algebra, where $l$ denotes the rank of the algebra.\footnote{We
recall that the rank is defined as the dimension of the Cartan subalgebras.} The eigenvalues of
Casimir operators are used
to label without ambiguity the irreducible representations of $\frak{g}$, while the states within a multiplet
can be distinguished using the generators of the Cartan subalgebra. In some situations, however, these
 operators are not enough to separate multiplicities, and following Racah \cite{Ra}, we need
 $f=\frac{1}{2}\left(\dim\frak{g}- 3l\right)$ additional operators (called Racah operators) to completely
classify states. The total number of internal labels required is thus
\begin{equation}
i=\frac{1}{2}(\dim \frak{g}- \mathcal{N}(\frak{g})).
\end{equation}

A similar situation holds whenever we use a some subalgebra
$\frak{h}$ to label the basis states of irreducible representations
of a Lie algebra $\frak{g}$.\footnote{This is what we will call ``missing
label problem'' (short MLP).} The subgroup
provides $\frac{1}{2}(\dim \frak{h}+\mathcal{N}(\frak{h}))+l^{\prime}$
labels, where $l^{\prime}$ denotes the number of invariants of
$\frak{g}$ that depend only on variables of the subalgebra
$\frak{h}$ \cite{Pe}. To separate states within irreducible
representations of $\frak{g}$ we thus need to find
\begin{equation}
n=\frac{1}{2}\left(
\dim\frak{g}-\mathcal{N}(\frak{g})-\dim\frak{h}-\mathcal{N}(\frak{h})\right)+l^{\prime}
\label{ML}
\end{equation}
additional operators, which we call missing label
operators. The total number of available operators of this kind is
easily shown to be twice the number of needed labels, i.e.,
$m=2n$. For $n>1$, it remains the problem of determining a set of
$n$ mutually commuting operators in order to prevent non-desired interactions.

\medskip
The analytical approach to the
missing label problem has the advantage of being formally very similar
to the problem of finding the generalized Casimir invariants of Lie algebras.
Although in general the missing label operators are neither invariants of the algebra nor any
of its subalgebras, they can actually be determined by means of differential equations with the same ansatz
as the general invariant problem \cite{Pe,Wi,C57,Ch,C67,BB,AA,Boy}.

\medskip

Given a Lie algebra $ \frak{g}=\left\{X_{1},..,X_{n}\; |\;
\left[X_{i},X_{j}\right]=C_{ij}^{k}X_{k}\right\}$ in terms of generators
and commutation relations, we
are primarily interested in (polynomial) operators
$C_{p}=\alpha^{i_{1}..i_{p}}X_{i_{1}}..X_{i_{p}}$ in the
generators of $\frak{s}$ such that the constraint $
\left[X_{i},C_{p}\right]=0$,\; ($i=1..n$) is satisfied. Such an
operator can be shown to lie in the centre of the enveloping algebra
of $\frak{s}$, and is traditionally referred to as Casimir
operator. However, in many dynamical problems, the relevant
invariant functions are not polynomials, but rational or even
transcendental functions (e.g. solvable groups in integrable systems or the inhomogeneous Weyl group).
Thus the approach with the universal enveloping algebra has
to be generalized in order to cover arbitrary Lie groups. The most
convenient method is the analytical realization. The generators
of the Lie algebra $\frak{s}$ are realized in the space
$C^{\infty }\left( \frak{g}^{\ast }\right) $ by means of the
differential operators:
\begin{equation}
\widehat{X}_{i}=C_{ij}^{k}x_{k}\frac{\partial }{\partial x_{j}},
\label{Rep1}
\end{equation}
where $\left\{ x_{1},..,x_{n}\right\}$ is a dual
basis of $\left\{X_{1},..,X_{n}\right\} $. The invariants of
$\frak{g}$ (in particular, the Casimir operators) are solutions of
the following system of partial differential equations:
\begin{equation}
\widehat{X}_{i}F=0,\quad 1\leq i\leq n.  \label{sys}
\end{equation}
Whenever we have a polynomial solution of (\ref{sys}), the
symmetrization map defined by
\begin{equation}
Sym(x_{i_{1}}^{a_{1}}..x_{i_{p}}^{a_{p}})=\frac{1}{p!}\sum_{\sigma\in
S_{p}}x_{\sigma(i_{1})}^{a_{1}}..x_{\sigma(i_{p})}^{a_{p}}
\end{equation}
allows to recover the Casimir operators in their usual form, i.e,
as elements in the centre of the enveloping algebra of $\frak{g}$.
A maximal set of functionally independent invariants is usually
called a fundamental basis. The number $\mathcal{N}(\frak{g})$ of
functionally independent solutions of (\ref{sys}) is obtained from
the classical criteria for differential equations, and is given
by:
\begin{equation}
\mathcal{N}(\frak{g}):=\dim \,\frak{g}- {\rm rank}\left(
C_{ij}^{k}x_{k}\right), \label{BB}
\end{equation}
where $A(\frak{g}):=\left(C_{ij}^{k}x_{k}\right)$ is the matrix
associated to the commutator table of $\frak{g}$ over the given
basis.

\medskip

If we now consider an algebra-subalgebra chain $\frak{s}\supset\frak{s}%
^{\prime}$ determined by an embedding $f:\frak{s}^{\prime}\rightarrow \frak{s}$,
in order to compute
the missing label operators we have to consider the equations of
(\ref{sys}) corresponding to the generators of the subalgebra
$\frak{s}^{\prime}$. This system, as proven in \cite{Pe}, has
exactly $\mathcal{N}(f(\frak{s}^{\prime}))=\dim \frak{s}-\dim
\frak{s}^{\prime}-l^{\prime}$ solutions. Using formula (\ref{ML})
it follows further that this scalar can be expressed in terms of
the number of invariants of the algebra-subalgebra chain:
\begin{equation}
\mathcal{N}(f(\frak{s}^{\prime}))=m+\mathcal{N}(\frak{s})+\mathcal{N}(
\frak{s}^{\prime})-l^{\prime}. \label{ML2}
\end{equation}
This shows that the differential equations corresponding to the
subalgebra generators have exactly $n$ more solutions as needed to
solve the missing label problem. The scalar $m$
depends essentially on how the subalgebra is embedded. In general,
to find a complete set of solutions for the labelling problem is a
difficult task. One of the main objectives is to find a sufficient
number of solutions without explicitly integrating the corresponding
system, but using the properties of the inclusion $\frak{s}^{\prime}\
subset\frak{s}$ and some related objects like contractions of Lie algebras.

\medskip

Contractions have developed formal formal procedure to justify
certain physical systems to a technique of considerable importance
\cite{Vi,IW,Lyk,We,C23}. It does not only allow to relate different
symmetry or classification schemes by means of limiting precesses,
but also provides useful information on the behavior of certain
observables and quantum numbers, codified in appropriate way by
invariant functions or Lagrangians. Various types of contractions
have been developed in the literature, and their equivalence or
relations have been explored. For the MLP that interests us, only
a quite specific type of contractions is of interest, that
corresponds to the symmetry breaking with respect to some inner
symmetry group. Therefore the presentation will be restricted to
this type of contractions: Let $\frak{g}$ be  a Lie algebra and
$\Phi_{t}\in End(\frak{g})$ a family of non-singular linear maps,
where $t\in [1,\infty)$.\footnote{Other authors use the parameter
range $t^{\prime}\in (0,1]$, which is equivalent to this by simply
changing the parameter to $t^{\prime}=1/t$.} For any
$X,Y\in\frak{g}$ we define
\begin{equation}
\left[X,Y\right]_{\Phi_{t}}:=\Phi_{t}^{-1}\left[\Phi_{t}(X),\Phi_{t}(Y)\right],
\end{equation}
which obviously represent the brackets of the Lie algebra over the
transformed basis, and defines an isomorphic algebra. Suppose that the limit
\begin{equation}
\left[X,Y\right]_{\infty}:=\lim_{t\rightarrow
\infty}\Phi_{t}^{-1}\left[\Phi_{t}(X),\Phi_{t}(Y)\right]
\label{Ko}
\end{equation}
exists for any $X,Y\in\frak{g}$. Then equation (\ref{Ko}) defines
a Lie algebra $\frak{g}^{\prime}$ called the contraction of
$\frak{g}$ (by $\Phi_{t}$), non-trivial if $\frak{g}$ and
$\frak{g}^{\prime}$ are non-isomorphic, and trivial otherwise
\cite{IW,We}. A contraction for which there exists some basis
$\left\{X_{1},..,X_{n}\right\}$ such that the contraction matrix
$A_{\Phi}$ is diagonal, that is, adopts the form
\begin{equation}
(A_{\Phi})_{ij}= \delta_{ij}t^{n_{j}},\quad n_{j}\in\mathbb{Z},
t>0, \label{IWK}
\end{equation}
is  called a generalized In\"on\"u-Wigner contraction \cite{We}.
This is the only type of contractions that we will need in this
work. Among the various properties of contractions, we enumerate
a numerical inequality satisfied by them that will play a central
role (for others see e.g. \cite{C23}): For an arbitrary contraction
$\frak{g}\rightsquigarrow \frak{g}^{\prime}$ the
following must hold:
\begin{equation}
\mathcal{N}\left( \frak{g}\right)  \leq\mathcal{N}\left(
\frak{g}^{\prime }\right). \label{KB}
\end{equation}

In analogy to the limiting process of the structure tensor, a notion of contraction
of invariants and Casimir operators can also be developed \cite{We2,WW3}.
The procedure is formally valid for polynomial and non-polynomial invariants, but we will
only consider Casimir operators here. Suppose that the contraction is
of the type (\ref{IWK}). If
$F(X_{1},...,X_{n})=\alpha^{i_{1}...i_{p}}X_{i_{1}}...X_{i_{p}}$
is a Casimir operator of degree $p$, then the transformed
invariant takes the form
\begin{equation}
F(\Phi_{t}(X_{1}),..,\Phi_{t}(X_{n}))=t^{n_{i_{1}}+...+n_{i_{p}}}\alpha^{i_{1}...i_{p}}X_{i_{1}}...X_{i_{p}}.
\end{equation}
Now, defining
\begin{equation}
M=\max \left\{n_{i_{1}}+...+n_{i_{p}}\quad |\quad
\alpha^{i_{1}..i_{p}}\neq 0\right\},
\end{equation}
the limit
\begin{equation}
F^{\prime}(X_{1},..,X_{n})=\lim_{t\rightarrow \infty}
t^{-M}F(\Phi_{t}(X_{1}),...,\Phi_{t}(X_{n}))=\sum_{n_{i_{1}}+...+n_{i_{p}}=M}
\alpha^{i_{1}...i_{p}}X_{i_{1}}...X_{i_{p}}
\end{equation}
gives a Casimir operator of degree $p$ of the contraction
$\frak{g}^{\prime}$. It should be remarked that, starting from an
adequate fundamental system of invariants
$\left\{C_{1},..,C_{p}\right\}$ of $\frak{g}$, it is always
possible to obtain a set of $p$ independent invariants of the
contraction. It should be observed that it is not ensured that these invariants are
of minimal degree in the contraction \cite{C72}, or even that they split into a sum
of more elementary invariants of the contraction.

\section{Embeddings and contractions of Lie algebras}

An embedding of a Lie algebra $\frak{s}^{\prime}$ into a Lie
algebra $\frak{s}$ is determined by an isomorphic mapping
$f:\frak{s}^{\prime}\longrightarrow \frak{s}$. In the case of
semisimple Lie algebras, the image of the subalgebra generators
can be described in terms of the usual Cartan-Weyl basis
$\left\{h_{k},e_{\alpha}\right\}$ of $\frak{s}$\footnote{$h_{k}$
denotes a generator in the Cartan subalgebra, while the
$e_{\alpha}$ correspond to the root vectors.} by:
\begin{equation*}
f(x)=\sum_{k=1}^{{\rm rank}\frak{s}}
a_{k}h_{k}+\sum_{\alpha\in\Delta} b_{\alpha}e_{\alpha},\quad x\in\frak{s}^{\prime}.
\end{equation*}
Embeddings are classified up to inner automorphisms of $\frak{s}$, and reduce
the classification to the determination of the non-equivalent embeddings classes. The question
that interest us is the behavior of representations of a simple Lie algebra when restricted to a
(semisimple) subalgebra. An important fact is that any embedding determines a integer factor $j_{f}$ given by
the relation
\begin{equation}
\left(f(x),f(x^{\prime}\right)=j_{f}\left(x,x^{\prime}\right),
\end{equation}
where $\left(.,.\right)$ is the usual scalar product defined on the dual Cartan subalgebras \cite{Serre}.
This scalar, being an invariant of the embedding class, constitutes a first label to distinguish reduction
chains. Generally we call this scalar the index of $\frak{s}^{\prime}$ in the Lie algebra $\frak{s}$. The index
has various important properties, from which we recall only those that will be of importance in the labelling
problem. At first, given disjoint subalgebras $\frak{s}_{j}^{\prime}$ of $\frak{s}$, the direct sum of the
subalgebras defines an embedding $f=\sum f_{i}$, the index of which is simply the sum of the various indices
$j_{f,i}$. Further, for reduction chains $\frak{s}\supset \frak{s}^{\prime}\supset\frak{s}^{\prime\prime}$,
the index of the last algebra in $\frak{s}$ is the product of the corresponding indices of the chain members.
The most important property used here concern the representations. Given $f:\frak{s}^{\prime}\rightarrow\frak{s}$
and a linear representation $\Phi$ of the latter algebra, then the indices of the representations
\footnote{We recall that the index $l_{\Phi}$ of a representation $R$ of highest weight $\Lambda$
is determined by the eigenvalue of the quadratic Casimir operator of the algebra multiplied by the factor
$\frac{\dim R}{\dim adj}$, where $adj$ denotes the adjoint representation \cite{Serre}.} are related by the formula:
\begin{equation}
j_{f}=\frac{l_{f\Phi}}{l_{\Phi}},
\end{equation}
where $l_{f\Phi}$ denotes the index of the induced representation on the subalgebra. We remark that this relation can be
useful for checking the existence of embeddings with a fixed branching rule.

\medskip

Among the different possibilities of embeddings, special types like regular
subalgebra, which can be directly obtained from the Dynkin
diagram of semisimple Lie algebras, or $S$-subalgebras, are of capital
importance in the theory of semisimple Lie algebras, and are well known \cite{Dy}.
The key fact is that the index $j_{f}$ serves to recognize the branching rules induced by
the embedding. Complete tables of branching
rules have not been obtained, although for simple
complex Lie algebras and maximal semisimple subalgebras, these
are tabulated up to rank eight \cite{Mc}.
As a special case of these branching rules, which is moreover the important case
for the missing label problem, a reduction chain
$\frak{s}^{\prime}\hookrightarrow_{f}\frak{s}$ determines the following
decomposition of the adjoint representation of $\frak{s}$:
\begin{equation}
{\rm ad} \frak{s}= {\rm ad} \frak{s}^{\prime}\oplus R.\label{D1}
\end{equation}
Here $R$ is a (completely reducible) representation of
$\frak{s}^{\prime}$ determined by the embedding index
$j_{f}$.\footnote{Complete reducibility is actually ensured
only if the subalgebra $\frak{s}^{\prime}$ is semisimple.} The latter
equation reflects a basis of $\frak{s}$ that is obtained starting from an
arbitrary basis of $\frak{s}^{\prime}$, and takes into account how the generators
of the Lie algebra are coupled with those of the subalgebra (which determines the
decomposition of the representation $R$ into IRs of the subalgebra).

\medskip

The crucial point is to construct the contraction related to the
reduction chain $\frak{s}^{\prime}\subset \frak{s}$. To this
extent, consider a basis $\left\{
X_{1},..,X_{s},X_{s+1},..,X_{n}\right\} $ of $\frak{s}$ such that
$\left\{  X_{1},..,X_{s}\right\}  $ is a basis of
$\frak{s}^{\prime}$, and $\left\{ X_{s+1},..,X_{n}\right\}  $
spans the representation space of the induced $R$. This basis adapted
to the subalgebra exactly reproduces the specific structure of the embedding.
The structure tensor of $\frak{s}$ can thus be rewritten as:
\begin{eqnarray}
\left[  X_{i},X_{j}\right]=&\sum_{k=1}^{s}C_{ij}^{k}X_{k},\;1\leq
i,j,k\leq s,\nonumber\\
\left[  X_{i},X_{j}\right]=&
\sum_{k=s+1}^{n}C_{ij}^{k}X_{k},\;1\leq i\leq
s,\;s+1\leq j,k \leq n,\nonumber\\
\left[  X_{i},X_{j}\right] =&\sum_{k=1}^{s}C_{ij}^{k}X_{k}+\sum
_{l=s+1}^{n}C_{ij}^{l}X_{l},\;s+1\leq i,j\leq n.
\end{eqnarray}
For any $t\in\mathbb{R}$ we consider the non-singular linear transformations%
\begin{equation}
\Phi_{t}\left(  X_{i}\right)  =\left\{
\begin{array}
[c]{cc}%
X_{i}, & 1\leq i\leq s\\
\frac{1}{t}X_{i}, & s+1\leq i\leq n
\end{array}
\right.  .\label{TB}
\end{equation}
Expressing the brackets over the transformed basis $\left\{
X_{i}^{\prime }=\Phi_{t}\left(  X_{i}\right)  :\;1\leq i\leq
n\right\}  $ we obtain
\begin{eqnarray}
\left[  X_{i}^{\prime},X_{j}^{\prime}\right]    & =\sum_{k=1}^{s}C_{ij}%
^{k}X_{k}^{\prime},\;1\leq i,j,k\leq s,\nonumber\\
\left[  X_{i}^{\prime},X_{j}^{\prime}\right]    & =\sum_{k=s+1}^{n}C_{ij}%
^{k}X_{k}^{\prime},\;1\leq i\leq s,\;s+1\leq j,k\leq n,\nonumber\\
\left[  X_{i}^{\prime},X_{j}^{\prime}\right]    &
=\sum_{k=1}^{s}\frac
{1}{t^{2}}C_{ij}^{k}X_{k}^{\prime}+\sum_{l=s+1}^{n}\frac{1}{t}C_{ij}^{l}%
X_{l}^{\prime},\;s+1\leq i,j\leq n.
\end{eqnarray}
It is straightforward to verify that the subalgebra $\frak{s}^{\prime}$ remains
invariant, as well as the representation $R$ of $\frak{s}^{\prime}$
over its complementary in $\frak{s}$. This is related to the fact that contractions
cannot modify branching rules, and therefore the type of Levi decompositions \cite{C72}.
These equations also show
that the limit
\begin{equation}
\lim_{t\rightarrow\infty}\Phi_{t}^{-1}\left[  \Phi_{t}\left(
X\right) ,\Phi_{t}\left(  Y\right)  \right]
\end{equation}
exists for any pair of generators $X,Y\in\frak{s}$, we thus obtain
a non-trivial contraction (actually a simple
In\"on\"u-Wigner contraction) of $\frak{s}$ denoted by $\frak{g}$ and with
non-vanishing brackets
\begin{eqnarray}
\left[  X_{i}^{\prime},X_{j}^{\prime}\right]   = &\sum_{k=1}^{s}C_{ij}%
^{k}X_{k}^{\prime},\;1\leq i,j,k\leq s,\nonumber\\
\left[  X_{i}^{\prime},X_{j}^{\prime}\right]   = & \sum_{k=s+1}^{n}C_{ij}%
^{k}X_{k}^{\prime},\;1\leq i\leq s,\;s+1\leq j,k\leq n.
\end{eqnarray}
We observe that if $\frak{s}^{\prime}$ is semisimple, then it
coincides
with the Levi subalgebra of $\frak{g}$, and the Levi decomposition of this contraction equals%
\begin{equation}
\frak{g}=\frak{s}^{\prime}\overrightarrow{\oplus}_{R}\left(
n-s\right) L_{1},
\end{equation}
where $(n-s)L_{1}$ denotes the Abelian algebra of dimension $n-s$.
This Lie algebra is affine, and by the contraction we know that
$\mathcal{N}(\frak{g})\geq \mathcal{N}(\frak{s})$. Applying the
analytical method, the invariants of $\frak{g}$ are obtained from
the solutions of the system:
\begin{eqnarray}
\widehat{X}_{i}F=C_{ij}^{k}x_{k}\frac{\partial F}{\partial
x_{j}}=0,& \quad 1\leq i\leq s,\nonumber \label{KS1}\\
\widehat{X}_{s+i}F=C_{s+i,j}^{s+k}x_{s+k}\frac{\partial
F}{\partial x_{j}}=0, & 1\leq i,k\leq n-s, 1\leq j\leq
s.\label{KS2}
\end{eqnarray}
The subsystem (\ref{KS1}) corresponds to the generators of
$\frak{s}^{\prime}$ realized as subalgebra of $\frak{s}$, while
the remaining equations (\ref{KS2}) describe the representation.
Written in matrix form, the system is given by
\[
\left(
\begin{array}
[l]{ccccll} 0 & ... & C_{1s}^{k}x_{k} &  C_{1,s+1}^{k}x_{k} & ...
&  C_{1,n}^{k}x_{k}\\
\vdots & & \vdots & \vdots & & \vdots\\
 -C_{1s}^{k}x_{k} & ... & 0 &  C_{s,s+1}^{k}x_{k} & ... &
 C_{s,n}^{k}x_{k}\\
 -C_{s,s+1}^{k}x_{k} & ... &  -C_{s,s+1}^{k}x_{k} &  0 &...&
 0\\
\vdots & & \vdots & \vdots & & \vdots\\
 -C_{1n}^{k}x_{k}& ...&  -C_{s,n}^{k}x_{k}&  0& ... & 0
\end{array}
\right)\left(
\begin{array}
[c]{c}
\partial_{x_{1}}F\\
\vdots\\
\partial_{x_{s}}F\\
\partial_{x_{s+1}}F\\
\vdots\\
\partial_{x_{n}}F\\
\end{array}\right)=0.
\]
Since the first $s$ first rows reproduce exactly the system
 of PDEs needed to compute the missing label operators, we
 conclude that any invariant of $ \frak{g}$ is a candidate for missing label
operator whenever it is functionally independent from the invariants of
$\frak{s}$ and $\frak{s}^{\prime}$.

\medskip

The following questions arise naturally from this ansatz:
\begin{enumerate}
\item  Do polynomial functions of the invariants of these algebras
suffice to determine $n$  mutually orthogonal  missing label
operators?

\item  Are all  available operators obtainable by this
procedure?
\end{enumerate}

Although the answer to both question is in the negative in the most general case,
it is in the affirmative for the first question for those reduction chains
for which the contraction provides a number of independent
invariants exceeding the number of needed labelling operators. It can
fails when these two quantities coincide, which suggests that the
procedure has to be refined. In some cases, the necessary operators cannot
be obtained from this refinement, and we have to develop additional machinery
to construct a set with the required operators. As concerns the second question, in general
there will be solutions that do not arise from the contraction and successive refinements,
 although it cannot be excluded that in some special cases we are able to recover a complete set of
independent labelling operators. As a general observation, only half of the available operators should be
expected, since all operators obtained are the result, in some
sense, of ``breaking" the original Casimir operators. This fact also suggest some ``inner'' hierarchy
of labelling operators, one of the classes corresponding to pure formal labelling operators without
an apparent physical meaning, in the sense that they cannot be obtained or deduced from the initial
data of the problem, and another class obtained as ``broken'' Casimir operators, which have a physical
meaning as the terms of the original invariants that remain preserved by the limit. This idea will be precised
more carefully later.

\medskip

In any case, for the contraction following inequality holds: $\mathcal{N}(f(\frak{s}^{\prime}))\geq
\mathcal{N}(\frak{g})$. Combining the latter with formula
(\ref{ML2}), we conclude that
\begin{equation}
\mathcal{N}(f(\frak{s}))=m+\mathcal{N}(\frak{s})+\mathcal{N}(\frak{s}^{\prime})-l^{\prime}\geq
\mathcal{N}(\frak{g})\geq \mathcal{N}(\frak{s}).\label{ML3}
\end{equation}
The term $\mathcal{N}(f(\frak{s}))$ on the left hand side gives
the total number of available labelling operators, the invariants
of $\frak{s}$ and $\frak{s}^{\prime}$ comprised, as shown in
\cite{Pe,Sh2}. Therefore, if the contraction $\frak{g}$ has enough
invariants, we can extract a set of $n$ commuting missing label
operators and solve the missing label problem completely.

Usually, we will be concerned with reduction
chains of the type $\frak{s}\supset\frak{s}^{\prime}$, where
$\frak{s}$ is semisimple and $\frak{s}^{\prime}$ is a reductive
Lie algebra. We remark that the contraction method remains completely
valid for reductions involving non-reductive algebra-subalgebra
chains. as a special type involving Cartan subalgebras, that turns out to be
of interest in labelling problems of semisimple Lie algebras of higher rank in
connection with spectroscopical applications, where they were first considered
\cite{Ra}. More recently, this unusual class of MLP has been considered in classification
schemes in chemical physics \cite{Ru,Ki}.

\section{MLPs solved with contractions only}

We begin analyzing the cases where the contraction $\frak{g}$ allows to solve the MLP
in satisfactory manner, and to set the limitations of this first approach. Some secondary
results will emerge, specially concerning bounds for the number of invariants in Lie algebras
arising by contraction. We assume that $\frak{s}$ is a semisimple Lie algebra of rank $p$,
$\frak{s}^{\prime}$ is a
reductive subalgebra and denote by $\frak{g=s}^{\prime}\overrightarrow{\oplus}%
_{R}(\dim\frak{s}-\dim\frak{s}^{\prime})kL_{1}$ the contraction associated to the chain $\frak{s}%
\supset\frak{s}^{\prime}$. Let $\left\{  C_{1},..,C_{p}\right\}  $
be the Casimir operators of $\frak{s}$, and $\left\{
D_{1},..,D_{q}\right\}  $ the invariants of $\frak{s}^{\prime}$.
Contracting the invariants $C_{i}$ or some appropriate combination
of them, we can always obtain $p$ independent invariants of
$\frak{g}$. Completing if necessary to a maximal set of invariants
of $\frak{g}$, we obtain the fundamental system  $\left\{
C_{1}^{\prime},..,C_{p}^{\prime},..,C_{r}^{\prime}\right\}
\;$($r\geq p$). In order to solve the missing label problem using
the latter\ set of functions, the system $\mathcal{F}=\left\{
C_{1}^{\prime },..,C_{r}^{\prime}\right\}  $  must contain at
least $n$ functions that are independent on the Casimir invariants
of $\frak{s}$ and $\frak{s}^{\prime}$, i.e.,
\begin{equation}
{\rm rank}\,\mathcal{F}\; \left(  {\rm mod} \left\{  C_{1},..,C_{p}%
,D_{1},..D_{q}\right\}  \right)  \geq n.\label{U1}%
\end{equation}
By construction, $\left\{
C_{1},..,C_{p},D_{1},..,D_{q-l^{\prime}}\right\}  $  are functionally independent.
 Now the question arises whether adding
the invariants of $\frak{g}$ some dependence relations appear. In
general, and whenever no invariant is preserved by the
contraction, the functions $C_{i}$ and $C_{i}^{\prime}$ are
independent. In this case a dependence relation means that some
$C_{i}$ is a function of $C_{i}^{\prime}$ and the invariants of
$\frak{s}^{\prime}$. Such dependence relations
appears for the quadratic Casimir operator
$C_{1}$.\footnote{Is either $\frak{s}$ or $\frak{s}^{\prime}$ is
not reductive, this is not applicable, since existence of
quadratic operators is not ensured.} Writing $C_{1}$ over
the transformed basis (\ref{TB}) we obtain the following
decomposition of $C_{1}$ as polynomial in the contraction variable
$t$:
\[
C_{1}=F+t^{2}C_{1}^{\prime},
\]
where $F$ is a quadratic invariant of $\frak{s}^{\prime}$. This
decomposition follows from the well known fact that, over the
given basis, the quadratic Casimir operator of a reductive
subalgebra is always a summand of the quadratic Casimir operator
of $\frak{s}$.\footnote{For higher order invariants, dependence
relations could also appear, depending on the homogeneity degree
of the invariants of $\frak{s}$ with respect to the generators of
the subalgebra.} As a consequence, we obtain the upper bound
\begin{equation}
{\rm rank}\left\{  C_{1},..,C_{p},C_{1}^{\prime},..,C_{r}^{\prime},D_{1}%
,..,D_{q}\right\}  <\mathcal{N}\left(  \frak{g}\right)
+\mathcal{N}\left( \frak{s}\right)  +\mathcal{N}\left(
\frak{s}^{\prime}\right)  -l^{\prime
}.\label{U2}%
\end{equation}
Combining the  bounds (\ref{U1}) and (\ref{U2})
respectively, we obtain a necessary numerical condition on the
number of invariants of the contraction $\frak{g}$:
\begin{equation}
n<\mathcal{N}\left(  \frak{g}\right).  \label{U3}%
\end{equation}

These facts, put together, allow us to decide when the
contraction $\frak{g}$ provides enough labelling operators to
solve the missing label problem for $\frak{s}\supset
\frak{s}^{\prime}$.

\begin{theorem}
A necessary and sufficient condition for solving the missing label
problem for the reduction $\frak{s}\supset\frak{s}^{\prime}$ by
means of the invariants of the associated contraction
$\frak{s\rightsquigarrow g=s}$ is that the affine Lie algebra
$\frak{g}$ satisfies the constraints

\begin{enumerate}
\item $\mathcal{N}\left(\frak{g}\right)  \geq n+1$,

\item  there are at least $n$ invariants of $\frak{g}$ that are
functionally independent from the invariants of $\frak{s}$ and
$\frak{s}^{\prime}$.
\end{enumerate}
\end{theorem}

The first condition, the easiest to evaluate, provides a numerical
criterion to decide whether the missing labels can be found by
means of the affine algebra $\frak{g}$. A sufficient condition can be obtained,
namely:

\begin{corollary}
If the contraction $\frak{g}$ satisfies the numerical condition
$\mathcal{N}(\frak{g})\geq \left\{n+1,
\mathcal{N}(\frak{s})+\mathcal{N}(\frak{s}^{\prime})+1-l^{\prime}\right\}$,
then it solves the MLP.
\end{corollary}

\medskip

Let
$\frak{s}^{\prime}\hookrightarrow_{f_{1}}\frak{s}$ be an embedding
and
$\frak{s\rightsquigarrow g}=\frak{s}^{\prime}\overrightarrow{\oplus}_{R}%
kL_{1}$ the associated contraction. The subalgebra
$\frak{s}^{\prime}$ is invariant by the contraction, we
naturally obtain the embedding
$f_{2}:\frak{s}^{\prime}\rightarrow\frak{g}$. Consider
the missing label problem for the latter
embedding.\footnote{Actually the mappings $f_{1}$ and $f_{2}$ are
the same, but we distinguish the target algebra by the indices.}
It follows that the system of PDEs to be solved is exactly
the same as for the embedding $f_{1}$. This means that the
solutions coincide, and, in particular, their number. Therefore
that \ $\mathcal{N}\left( f_{1}\left( \frak{s}^{\prime}\right)
\right) =\mathcal{N}\left( f_{2}\left( \frak{s}^{\prime}\right)
\right) $. Recall that for each embedding the number of
independent solutions is given by
\begin{eqnarray}
\mathcal{N}\left(  f_{1}\left(  \frak{s}^{\prime}\right)  \right)
&
=\dim\frak{s}-\dim\frak{s}^{\prime}+l^{\prime},\nonumber \\
\mathcal{N}\left(  f_{2}\left(  \frak{s}^{\prime}\right)  \right)
& =\dim\frak{g}-\dim\frak{s}^{\prime}+l_{1}^{\prime}, \label{U4}
\end{eqnarray}
where $l_{1}^{\prime}$ denotes the number of common invariants of
$\frak{s}^{\prime}$ and $\frak{g}$. Since contractions preserve
the dimension, we conclude from formula (\ref{U4}) that
$l^{\prime}=l_{1}^{\prime}$, that is, the subalgebra
$\frak{s}^{\prime}$ has the same number of common invariants with
$\frak{s}$ than with the contraction $\frak{g}$. On the other
hand, using the reformulation (\ref{ML2})
\begin{eqnarray}
\mathcal{N}\left(  f_{1}\left(  \frak{s}^{\prime}\right)  \right)
& =m+\mathcal{N}\left(  \frak{s}\right)  +\mathcal{N}\left(
\frak{s}^{\prime
}\right)  -l^{\prime}\nonumber \\
\mathcal{N}\left(  f_{2}\left(  \frak{s}^{\prime}\right)  \right)
& =\widetilde{m}+\mathcal{N}\left(  \frak{g}\right)
+\mathcal{N}\left(
\frak{s}^{\prime}\right)  -l_{1}^{\prime}%
\end{eqnarray}
we deduce that
\begin{equation}
m-\widetilde{m}=\mathcal{N}\left(  \frak{g}\right)
-\mathcal{N}\left( s\right)  \geq0.\label{ML3}
\end{equation}
This result tells us that the number of available labelling
operators for the reduction chain
$\frak{s}\supset\frak{s}^{\prime}$ is always higher than that of
the chain $\frak{g}\supset\frak{s}^{\prime}$. From this we obtain
an interesting relation between the number of available operators
for the different embeddings $f_{1}$ and $f_{2}$: Let $\frak{s}\rightsquigarrow \frak{g}$
be such that
the subalgebra $\frak{s}^{\prime}$ is (maximal) invariant. Then
following equality holds:
\[
\mathcal{N}\left( \frak{g}\right)=\mathcal{N}\left(
s\right)+m-\widetilde{m},
\]
where $m$ and $\widetilde{m}$ is the number of available missing
label operators for the algebra subalgebra chain $\frak{s}\supset
\frak{s}^{\prime}$ and $\frak{g}\supset \frak{s}^{\prime}$,
respectively.

As special case, we get the following upper bound
\begin{equation}
\mathcal{N}\left( \frak{g}\right)\leq \mathcal{N}\left(
s\right)+m. \label{ML5}
\end{equation}
This bound points out  that the
number of invariants of a contraction is, in some sense,
determined by the number of available missing label operators for
the missing label problem with respect to a maximal subalgebra of
$\frak{s}$ that remains invariant by the contraction. Observe that the
essential vanishing of brackets occurs in the maximal solvable ideal, since
the subalgebra and the branching rule remains fixed.

\medskip

Thus, for low values of $n$ the contraction is an effective tool to solve the MLP,
as well as for cases with a large number of invariants for the contracted Lie algebra $\frak{g}^{\prime}$.
We review some of these cases with their most representative physical chains.

\subsection{The case $n=m=0$}

In the case of zero missing labels, the invariants of the
algebra-subalgebra chain provide a complete description of the
states. This situation is not uncommon for certain canonical
embeddings, such as the inclusions $\frak{so}(p,q)\subset
\frak{so}(p,q+1)$ of (pseudo)-orthogonal Lie algebras. Even if this
case is trivial, its interpretation in terms of the associated
contraction provides some interesting information concerning the
invariants of the contraction.

If $m=0$, then formula (\ref{ML3}) implies that
$\mathcal{N}\left(  \frak{g}\right)=\mathcal{N}\left( s\right)$,
that is, the contraction associated to the embedding
$\frak{s}\supset\frak{s}^{\prime}$ preserves the number of
invariants (the converse does not
necessarily hold). Moreover, by formula (\ref{ML}), we have
\begin{equation}
0=m=\dim \frak{s}-\dim
\frak{s}^{\prime}-\mathcal{N}(\frak{s})-\mathcal{N}(\frak{s}^{\prime})+2l^{\prime}.
\end{equation}
In absence of additional internal labels, the system
$\widehat{X}_{i}F=0$ for the generators of $\frak{s}^{\prime}$ has
exactly
\begin{equation}
\mathcal{N}(f(\frak{s}))=\mathcal{N}(\frak{s})+\mathcal{N}(\frak{s}^{\prime})-l^{\prime}
\end{equation}
solutions. Since any invariant of the contraction $\frak{g}=\frak{s}^{\prime}\overrightarrow{\oplus}_{R}%
(\dim\frak{s}-\dim\frak{s}^{\prime})L_{1}$ is a special solution
of this system, the latter equation tells that any invariant of
$\frak{g}$ is functionally dependent on the invariants of
$\frak{s}$ and the subalgebra $\frak{s}^{\prime}$. That is, the
Casimir invariants of the algebra-subalgebra chain completely
determine the invariants of the contraction.\footnote{Of course,
if $\mathcal{N}(\frak{s}^{\prime})=0$, this assertion fails, but
for reductive subalgebras this situation is excluded.} Expressed
in another way, in this situation, polynomial functions of the
invariants of $\frak{s}$ and the contraction $\frak{g}$ allow to
recover naturally the invariants of the subalgebra.

\medskip

Typical chains where the number of labelling operators is zero are the
pseudo-orthogonal reductions $\frak{so}(p,q)\supset\frak{so}(p-1,q)$
and $\frak{so}(p,q)\supset\frak{so}(p,q-1)$. This has been used to analyze the
corresponding inhomogeneous algebras \cite{C49}, and justifies to some extent
the validity of the Gel'fand method for non-semisimple Lie algebras. Another
interesting class of algebras where $m=0$ holds is the extended Schr\"odinger algebra,
$\widehat{S}(N)$, which is the invariance algebra of the Schr\"odinger equation in $(N+1)$-
dimensional spacetime. The remarkable fact is that this algebra is no more semisimple, but
a semidirect product of a semisimple algebra with a Heisenberg-Weyl algebra \cite{C46}.

\subsection{The case $n=1, m=2$}

For the case of one missing label operator, any solution of the
contraction $\frak{g}$ that is independent of the invariants of
the algebra-subalgebra chain is an admissible labelling operator.
Formula (\ref{ML5}) establishes the maximal
possible number for the invariants of $\frak{g}$:
\[
\mathcal{N}\left( \frak{g}\right)\leq \mathcal{N}\left(
s\right)+2.
\]
Observe that two is exactly the number of available operators.
There are eight cases with one missing label \cite{Pe,Lo},
semisimple Lie algebra $\frak{s}$ and maximal
reductive subalgebra $\frak{s}^{\prime}$. Most of these chains have
been solved explicitly finding finite integrity bases, that is, a set of
elementary subgroup scalar such that any other can be expressed by
a polynomial in them. All these can also be solved applying
the contraction method. The eight possibilities are resumed in Table 1.\footnote{This
enumeration should be understood in a broad sense. The number of labelling operators
does not depend on the embedding class, thus various different reduction chains are
considered as one possibility. Further, the different real forms of the algebras also
give rise to different MLPs, although the type is still the same.}

\subsection{The $\frak{su}\left(  3\right)  \supset\frak{so}\left(  3\right)
$ reduction}

This reduction, also called the Elliott chain, was introduced in order to generalize the
 group theoretical analysis developed for $L-S$ and $j-j$ coupling schemes to the mixing
 of two different orbital shells \cite{El}. This case is without doubt the best studied
 missing label problem. A complete set of commuting
operators and their eigenvalues for different irreducible
representations of $\frak{su}(3)$ were determined in
\cite{Za}.\newline The $\frak{so}\left(  3\right)  $ subalgebra is
naturally identified with the three orbital angular momentum
operators, while the remaining five generators transform under
rotations like the elements of a second rank tensor \cite{El,Ro2}.
Here we consider a basis $\left\{  L_{i},T_{jk}\right\}  $
formed by rotations $L_{i}$ and the operators $T_{ik}$ and commutation relations%
\begin{equation*}
\begin{tabular}
[c]{ll}%
$\left[  L_{j},L_{k}\right]  =i\varepsilon_{jkl}L_{l},$ & $\left[
L_{j},T_{kl}\right]  =i\varepsilon_{jkm}T_{lm}+i\varepsilon_{jlm}T_{km},$\\
\multicolumn{2}{l}{$\left[  T_{jk},T_{lm}\right]
=\frac{i}{4}\left\{
\delta_{j}^{l}\varepsilon_{kmn}+\delta_{j}^{m}\varepsilon_{k\ln}+\delta
_{k}^{l}\varepsilon_{jmn}+\delta_{k}^{m}\varepsilon_{j\ln}\right\}  L_{n},$}%
\end{tabular}
\end{equation*}
where $T_{33}+\left(  T_{11}+T_{22}\right)  =0$. The symmetrized
Casimir operators, following the notation of \cite{Za}, are given
by $C^{\left(
2\right)  }=L_{i}L_{i}+2T_{ik}T_{ik}$, $\ C^{\left(  3\right)  }=L_{i}%
T_{ik}L_{k}-\frac{4}{3}T_{ik}T_{kl}T_{li}$ and $C^{\left(  2,0\right)  }%
=L_{i}L_{i}$. The contraction $\frak{g}$ associated to this
reduction has Levi
decomposition $\frak{g=so}\left(  3\right)  \overrightarrow{\oplus}_{R_{5}%
^{I}}5L_{1}$, where $R_{5}^{I}$ denotes the five dimensional
irreducible representation of $\frak{so}\left(  3\right)  $. This
 is equivalent to the rotor algebra $[\mathbb{R}^{5}]SO(3)$
 studied in \cite{Ro}. It is straightforward to verify that $\mathcal{N}\left(
\frak{g}\right) =2$. Therefore, a basis of invariants of
$\frak{g}$ can be obtained by contraction of $C^{\left( 2\right)
}$ and $C^{\left( 3\right)  }$. Specifically, we get the
(unsymmetrized) Casimir invariants%
\begin{eqnarray*}
C_{2}  =2t_{ik}t^{ik},\\
C_{3}  =t_{ik}t^{kl}t_{li}.
\end{eqnarray*}
As already observed, $C_{2}$ is functionally dependent on
$C^{\left( 2\right) }$ and $C^{\left(  2,0\right)  }$, therefore
of no use for the MLP. The independence of $\left\{  C^{\left(
2\right) },C^{\left( 3\right)  },C^{\left(
2,0\right)  },C_{3}\right\}  $ follows from the Jacobian%
\[
\frac{\partial\left\{  C^{\left(  2\right)  },C^{\left(  3\right)
},C^{\left(  2,0\right)  },C_{3}\right\}  }{\partial\left\{  l_{2}%
,l_{3},t_{11},t_{12}\right\}  }\neq0.
\]
The invariant $C_{3}$ is therefore sufficient to solve the missing
label problem. In fact, we can recover the missing label operator
$X^{\left( 3\right)  }$ from \cite{Za} by simply considering the
linear combination
\[
X^{\left(  3\right)  }=C^{\left(  3\right)  }+\frac{4}{3}\left\{
C_{3}\right\}  _{symmetrized}.
\]
This operator is equivalent to the third order operator obtained
by Bargmann and Moshinsky in \cite{Ba}, and also to the operator
determined in \cite{Ro} using the K-matrix approach. It is
observed that the fourth order operator $X^{\left( 4\right)
}=L_{i}T_{ij}T_{jk}L_{k}$ cannot be obtained from the invariants
of $\frak{su}\left( 3\right) ,\frak{so}\left( 3\right)  $ and the
contraction $\frak{g}$. This is essentially due to the fact that
the fundamental Casimir operators of $\frak{su}(3)$ have degree
two and three. Another recent approach to this reduction chain can be
found in \cite{Jar}.

\section{The exceptional chain $G_{2}\supset \frak{su}(3)$}

The exceptional Lie group $G_{2}$ has been shown to have interesting physical applications,
as followed from Racah's on atomic spectroscopy, where it is essential for the understanding
of the $f=3$ shell \cite{Ra,Wy}. This group was also considered as candidate to describe strong
interactions, prior to the success of the unitary group $\frak{su}(3)$, as well as in the development of
an eight-fold way for the electronic $f$-shell \cite{Ju2,Wy2}. The fact that the latter
is contained as maximal subgroup in $G_{2}$ has made the exceptional group an interesting object in
hadron spectroscopy. For these applications, it is convenient to express $G_{2}$ in a $\frak{su}(3)$ basis.
It is easy to see the adjoint representation $\Gamma_{\left(  1,0\right)  }$
of $G_{2}$ decomposes like follows with respect to $\frak{su}(3)$:
\begin{equation}
\Gamma_{\left(  1,0\right)  }=8+3+\overline{3}%
\end{equation}
According to this decomposition, we label the generators as $E_{ij}%
,a_{k},b^{l}$ $\left(  i,j,k,l=1,2,3\right)$ (with the constraint $E_{11}+E_{22}+E_{33}=0$). We have the brackets:%
\begin{equation}
\begin{array}{ll}
\left[  E_{ij},E_{kl}\right] =   & \delta_{jk}E_{il}-\delta_{il}E_{kj}\\
\left[  E_{ij},a_{k}\right]    & =\delta_{jk}a_{i}\\
\left[  E_{ij},b^{k}\right]   = & -\delta_{ik}b^{j}\\
\left[  a_{i},a_{j}\right]   = & -2\varepsilon_{ijk}b^{k}\\
\left[  b^{i},b^{j}\right]   = & 2\varepsilon_{ijk}a_{k}\\
\left[  a_{i},b^{j}\right]   = & 3E_{ij}
\end{array}
\end{equation}
The subalgebra is clearly spanned by the operators $E_{ij}$. We moreover
 choose the Cartan subalgebra generated by the
operators $H_{1}=E_{11}-2E_{22}+E_{33}$ and $H_{2}=E_{22}-E_{33}$. The
operators $\left\{  a_{1},a_{2},a_{3}\right\}  $ correspond to the fundamental
quark representation $3$, while $\left\{  b^{1},b^{2},b^{3}\right\}  $ corresponds
to the antiquark representation $\overline{3}$.\footnote{The corresponding brackets
for the adjoint representation are given in Table 1.} Considering the reduction chain
$G_{2}\supset \frak{su}(3)$, we see that separation of multiplicities requires
\[
n=\frac{1}{2}\left(14-2-8-2\right)=1
\]
additional labelling operator.

\begin{table}
\caption{\label{t3}The adjoint representation of $G_{2}$ in the $A_{2}$-basis}
\begin{tabular*}{\textwidth}{@{}c|*{15}{@{\extracolsep{0pt plus
11pt}}c}}
& $V_{1}$ & $V_{2}$ & $V_{3}$ & $V_{4}$ & $V_{5}$ & $V_{6}$ & $V_{7}$ & $V_{8}$ & $V_{9}$ &$V_{10}$ & $V_{11}$ & $V_{12}$ & $V_{13}$ & $V_{14}$ \\\hline
$H_{1}$ & 0 & 0 & $2V_{3}$ & $-2V_{4}$ & $-3V_{5}$ & $3V_{6}$ & $-V_{7}$ & $V_{8}$ &
$V_{9}$ & $-V_{10}$ & $3V_{11}$ & $-3V_{12}$ & $0$ & $0$ \\
$H_{2}$ & 0 & 0 & $-V_{3}$ & $V_{4}$ & $2V_{5}$ & $-2V_{6}$ & $V_{7}$ & $-V_{8}$ &
$0$ & $0$ & $-V_{11}$ & $V_{12}$ & $V_{13}$ & $-V_{14}$ \\
$E_{12}$ & $-3V_{11}$ & $V_{11}$ & $0$ & $V_{9}$ & $V_{13}$ & $0$ & $0$ & $0$ & $0$ & $-V_{3}$ & $0$ & $W_{1}$ & $0$ & $-V_{6}$ \\
$E_{21}$ & $3V_{12}$ & $-V_{12}$ & $-V_{10}$ & $0$ & $0$ & $-V_{14}$ & $0$ & $0$ & $V_{4}$ & $0$ & $-W_{1}$ & $0$ & $V_{5}$ & $0$ \\
$E_{23}$ & $3V_{5}$ & $-2V_{5}$ & $-V_{7}$ & $0$ & $0$ & $V_{2}$ & $0$ & $V_{4}$ & $0$ & $0$ & $-V_{13}$ & $0$ & $0$ & $V_{12}$ \\
$E_{32}$ & $-3V_{6}$ & $2V_{6}$ & $0$ & $V_{8}$ & $-V_{2}$ & $0$ & $-V_{3}$ & $0$ & $0$ & $0$ & $0$ & $V_{14}$ & $-V_{11}$ & $0$ \\
$E_{13}$ & $0$ & $-V_{13}$ & $0$ & $0$ & $0$ & $V_{11}$ & $0$ & $V_{9}$ & $0$ & $-V_{7}$ & $0$ & $-V_{5}$ & $0$ & $W_{2}$ \\
$E_{31}$ & $0$ & $V_{14}$ & $0$ & $0$ & $-V_{12}$ & $0$ & $-V_{10}$ & $0$ & $V_{8}$ & $0$ & $V_{6}$ & $0$ & $-W_{2}$ & $0$ \\
$a_{1}$ & $-3V_{9}$ & $0$ & $3V_{11}$ & $-2V_{7}$ & $0$ & $0$ & $3V_{13}$ & $2V_{3}$ & $0$ & $W_{3}$ & $0$ & $-V_{4}$ & $0$ & $-V_{8}$ \\
$a_{2}$ & $2V_{4}$ & $-V_{4}$ & $-V_{1}$ & $0$ & $0$ & $-V_{8}$ & $3V_{5}$ & $-2V_{10}$ & $2V_{7}$ & $3V_{12}$ & $-V_{9}$ & $0$ & $0$ & $0$ \\
$a_{3}$ & $-3V_{8}$ & $V_{8}$ & $3V_{6}$ & $2V_{10}$ & $-V_{4}$ & $0$ & $-W_{4}$ & $0$ & $-2V_{3}$ & $3V_{14}$ & $0$ & $0$ & $-V_{9}$ & $0$ \\
$b^{1}$ & $V_{10}$ & $0$ & $2V_{8}$ & $-3V_{12}$ & $0$ & $0$ & $-2V_{4}$ & $-3V_{14}$ & $-W_{3}$ & $0$ & $V_{3}$ & $0$ & $V_{7}$ & $0$ \\
$b^{2}$ & $-2V_{3}$ & $V_{3}$ & $0$ & $V_{1}$ & $V_{7}$ & $0$ & $2V_{9}$ & $-3V_{6}$ & $-3V_{11}$ & $-2V_{8}$ & $0$ & $V_{10}$ & $0$ & $0$ \\
$b^{3}$ & $V_{7}$ & $-V_{7}$ & $-2V_{9}$ & $-3V_{5}$ & $0$ & $V_{3}$ & $0$ & $W_{4}$ & $-3V_{13}$ & $V_{4}$ & $0$ & $0$ & $0$ & $V_{10}$ \\\hline
\multispan{14}  $W_{1}=V_{1}+V_{2}, W_{2}=V_{1}+2V_{2}, W_{3}=2V_{1}+3V_{2}, W_{4}=V_{1}+3V_{2}$ {\hfill}&\\
\end{tabular*}
\end{table}

The contraction associated to the reduction $G_{2}\supset \frak{su}(3)$ has at least two
Casimir operators, thus the preceding results apply and the missing label problem can be
solved. As expected, the quadratic invariant of $G_{2}$ can be neglected for providing no
information. Following the procedure developed in \cite{Bi94,C42}, the sixth order Casimir operator is
rewritten as\footnote{The invariant has 432 terms over this basis, we therefore skip its explicit
expression here.}
\begin{equation}
C_{6}=3C_{\left[  2,4\right]  ,1}-6C_{\left[  4,2\right]  ,1}-\frac{3}%
{4}C_{\left[  2,4\right]  ,2}-9C_{\left[  2,4\right]  ,3}+9C_{\left[
4,2\right]  ,2}+27C_{\left[  3,3\right]  },
\end{equation}
where the $C_{\left[  i,j\right]  ,k}$ denote operators of degree $i$ in the
$\frak{su}\left(  3\right)  $ generators and degree $j$ in the representation
space variables, $k$ being an additional index to separate operators of the
same degree. It follows at once that $3C_{\left[  2,4\right]  ,1}-\frac{3}%
{4}C_{\left[  2,4\right]  ,2}-9C_{\left[  2,4\right]  ,3}$ is the Casimir
operator of the corresponding contraction, for having the highest power in the
variables of the representation of $\frak{su}\left(  3\right)  $. It can be
verified that the operator $\Omega=9C_{\left[  4,2\right]  ,2}+27C_{\left[
3,3\right]  }-6C_{\left[  4,2\right]  ,1}$ is independent on the Casimir
operators of both $G_{2}$ and $\frak{su}\left(  3\right)  $. In the absence of
orthogonality conditions, $\Omega$ provides a labelling operator for the
reduction.

\subsection{The $\frak{so}\left(  5\right)  \supset\frak{su}\left(
2\right) \times\frak{u}\left(  1\right)  $ chain}

The study of this reduction, also called the seniority model, was motivated
by the close connection between the Wigner coefficients involving the standard
representation of $\frak{so}(5)$ with the fractional parentage coefficients
 of spin-2 systems in the seniority scheme \cite{Ra,He,Hel}.

\medskip

To analyze this case, we consider the same basis $\left\{
U_{\pm},U_{3},V_{3},V_{\pm},S_{\pm },T_{\pm}\right\}  $ used in
\cite{Sh3,Gr}. The $\frak{su}\left(  2\right) \times\frak{u}\left(
1\right)  $ subalgebra is generated by the operators $\left\{
U_{\pm},U_{3},V_{3}\right\}  $. The nonzero brackets are given by
\begin{equation*}
\begin{tabular}
[c]{llll}%
$\left[  U_{\pm},U_{3}\right]  =\mp U_{\pm},$ & $\left[
U_{+},U_{-}\right] =2U_{3},$ & $\left[  U_{\pm},V_{\pm}\right]
=\mp2S_{\pm},$ & $\left[  U_{\pm
},V_{\mp}\right]  =\mp2T_{\pm},$\\
$\left[  U_{\pm},S_{\mp}\right]  =\pm V_{\mp},$ & $\left[
U_{\pm},T_{\mp }\right]  =\pm V_{\mp},$ & $\left[
U_{3},S_{\pm}\right]  =\pm S_{\pm},$ &
$\left[  U_{3},T_{\pm}\right]  =\pm T_{\pm},$\\
$\left[  V_{3},S_{\pm}\right]  =\pm S_{\pm},$ & $\left[
V_{3},T_{\pm}\right] =\mp T_{\pm},$ & $\left[  V_{+},V_{-}\right]
=2V_{3},$ & $\left[  V_{\pm
},V_{3}\right]  =\mp V_{\pm},$\\
$\left[  V_{\pm},S_{\mp}\right]  =\mp U_{\mp},$ & $\left[
V_{\pm},T_{\pm }\right]  =\pm U_{\pm},$ & $\left[
S_{+},S_{-}\right]  =U_{3}+V_{3},$ &
$\left[  T_{+},T_{-}\right]  =U_{3}-V_{3}.$%
\end{tabular}
\end{equation*}
Using standard methods, the (unsymmetrized) Casimir operators of
$\frak{so}\left( 5\right)  $ can be chosen as
\[%
\begin{tabular}
[c]{ll}%
$C_{2}=$ & $u_{+}u_{-}+u_{3}^{2}+v_{3}^{2}+v_{+}v_{-}+2\left(  s_{+}%
s_{-}+t_{+}t_{-}\right)  ,$\\
$C_{4}=$ & $\left(  u_{+}u_{-}+u_{3}^{2}\right)  v_{3}^{2}+u_{+}u_{-}\left(
s_{+}s_{-}+t_{+}t_{-}\right)  +u_{+}^{2}s_{-}t_{-}+u_{-}^{2}s_{+}t_{+}%
+2u_{3}v_{3}s_{+}s_{-}$\\
& $+\left(  \left(  t_{-}v_{-}-s_{-}v_{+}\right)  u_{+}+\left(  t_{+}%
v_{+}-s_{+}v_{-}\right)  u_{-}\right)  v_{3}+\left(  t_{+}v_{+}+s_{+}%
v_{-}\right)  u_{-}$\\
& $+v_{+}v_{-}s_{+}s_{-}+u_{3}^{2}v_{+}v_{-}+\left(  s_{+}s_{-}-t_{+}%
t_{-}\right)  ^{2}-v_{+}^{2}s_{-}t_{+}-v_{-}^{2}s_{+}t_{-}+v_{+}v_{-}%
t_{+}t_{-}$\\
& $+s_{-}v_{+}u_{+}u_{3}+t_{-}v_{-}u_{+}u_{3}-2u_{3}v_{3}t_{+}t_{-}  $.
\end{tabular}
\]

Those of the subalgebra are very easy to find:
$C_{21}=u_{+}u_{-}+u_{3}^{2}$ and $C_{22}=v_{3}$. In this case, the associated
contraction $\frak{g}$ is inhomogeneous with Levi part isomorphic to $\frak{su}(2)$,
and it preserves the number of invariants. Contracting the operators above, we get the
invariants:
\begin{equation}
\begin{array}{ll}
C_{2}^{\prime}  & =v_{+}v_{-}+2\left(  s_{+}s_{-}+t_{+}t_{-}\right)  ,\\
C_{4}^{\prime}  & =v_{+}v_{-}s_{+}s_{-}+\left(
s_{+}s_{-}-t_{+}t_{-}\right)
^{2}-v_{+}^{2}s_{-}t_{+}-v_{-}^{2}s_{+}t_{-}+v_{+}v_{-}t_{+}t_{-}.
\end{array}
\end{equation}
As expected, the quadratic Casimir operator does not provide useful
operators. Thus only $C_{4}^{\prime}$ can be used. In order to check
the independence of the latter from the Casimir operators of $\frak{so}(5)$
and the subalgebra, we compute the Jacobian
\[
\frac{\partial\left\{C_{2},C_{4},C_{4}^{\prime},C_{21},C_{22}\right\}}{\partial\left
\{u_{+},u_{-},v_{3},v_{+},v_{-}\right\}}\neq 0,
\]
that shows the possibility of solving the missing label problem for this chain.
Some manipulation of the preceding functions leads us to the labelling operator
$\Omega_{4}=C_{4}-C_{4}^{\prime}-C_{21}C_{22}^{2}$ given explicitly
by
\[
\begin{array}{ll}
\Omega_{4}=&u_{+}u_{-}\left(  s_{+}s_{-}+t_{+}t_{-}\right)  +u_{3}^{2}%
v_{+}v_{-}+u_{+}^{2}s_{-}t_{-}+u_{-}^{2}s_{+}t_{+}+2u_{3}v_{3}\left(
s_{+}s_{-}-t_{+}t_{-}\right) \\
&+\left(  \left(  t_{-}v_{-}-s_{-}v_{+}\right)  u_{+}+\left(  t_{+}v_{+}%
-s_{+}v_{-}\right)  u_{-}\right)  v_{3}+\left(  t_{+}v_{+}+s_{+}%
v_{-}\right)u_{-}u_{3}\\
& +  s_{-}v_{+}u_{+}u_{3}+t_{-}v_{-}u_{+}u_{3}.
\end{array}
\]
Symmetrizing this operator $\Omega_{4}$, we conclude that it coincides
 with the fourth order operator
$UVL^{2}$ constructed in \cite{Pe}. Since $m=2$, there is another possibility
for the labelling operator, of degree three. This cannot however be recovered
by the procedure, since odd Casimir operators do not exist for
the orthogonal algebra $\frak{so}(5)$.

\begin{table}
\caption{\label{Tab2} Missing label operators through contraction }
\begin{tabular}[c]{@{}l|cccc} $\frak{s\supset s}^{\prime}$
& $\mathcal{N}\left(  \frak{g}\right)  $ & $\mathcal{N}\left(
f\left(  \frak{s}^{\prime}\right)  \right)  $ & rank $\mathcal{F}$
& Order of $\Phi$ \\\hline
$\frak{su}\left(
3\right)  \supset\frak{so}\left(  3\right)  $ & $2$ & $5$ &
$4$ & $3$ \\
$\frak{so}\left(  5\right)  \supset\frak{su}\left(  2\right)
\times
\frak{u}\left(  1\right)  $ & $2$ & $6$ & $5$ & $4$ \\
$G_{2}\supset\frak{su}\left(  3\right)  $ & $2$ & $5$ & $4$ & $6$ \\
$\frak{sp}\left(  6\right)  \supset\frak{sp}\left(  4\right)
\times
\frak{su}\left(  2\right)  $ & $3$ & $8$ & $7$ & $6$ \\
$\frak{so}\left(  7\right)  \supset G_{2}$ & $3$ & $7$ & $6$ & $6$
\\
$\frak{su}\left(  4\right)  \supset\left[  \frak{su}\left(
2\right)  \right]
^{2}\times\frak{u}\left(  1\right)  $ & $3$ & $7$ & $6$ & $4$ \\
$\frak{su}\left(  3\right)  \times\frak{su}\left(  3\right)
\supset
\frak{su}\left(  3\right)  $ & $2$ & $8$ & $7$ & $3$ \\
$\left[\frak{su}\left(  2\right)\right]^{3}\supset\frak{su}\left(
2\right) $ & $3$ & $6$ & $5$ & $2$   \\\hline
\end{tabular}
\end{table}

\section{Reductions with $n>1$, $m>2$}

Reduction chains with more than two missing labels are notably more complicated,
mainly because of the requirement that the labelling operators found
 must commute. Many physically important cases belonging to this type have been analyzed in the
literature, although only for a small number the complete solution has been found. The best known
example is the Wigner supermultiplet $\frak{su}(4)\supset
\frak{su}(2)\times\frak{su}(2)$, studied algebraically by various authors, and for which the numerical
values of the labelling operators have been computed for large classes of irreducible representations
\cite{Que,Pa,Mo}. In this section we show that the approach of using only the invariants of the contraction associated
to the reduction holds for more than one labelling operator, and sometimes coincide with those operators
found by different procedures.

\subsection{ The supermultiplet model}

This reduction was considered to describe light nuclei, in opposition to
the isospin-strange spin contents of $\frak{su}(4)$, which uses the canonical
embedding of $\frak{su}(2)\times\frak{su}(2)$ into the Lie algebra. For the
multiplet model, the
set of available operators is partitioned into two separate sets,
the Moshinky-Nagel operators ${\Omega,\Phi}$ and two other
operators ${O_{1},O_{2}}$ found in \cite{Que}. An approach using contractions
can be found in \cite{C72}. We resume that result. Using the same basis
$\left\{ S_{i},T_{j},Q_{\alpha\beta}\right\}  $ of \cite{Mo},
where $1\leq i,j,\alpha,\beta\leq3$, the non-vanishing
brackets of $\frak{su}\left(  4\right)  $ are%
\begin{eqnarray}
\left[  S_{i},S_{j}\right]  =i\varepsilon_{ijk}S_{k},\; \left[  T_{i}%
,T_{j}\right]  =i\varepsilon_{ijk}T_{k},\; \left[
S_{i},Q_{j\alpha}\right] =i\varepsilon_{ijk}Q_{k\alpha},\; \left[
T_{\alpha},Q_{i\beta}\right]
=i\varepsilon_{\alpha\beta\gamma}Q_{i\gamma},\nonumber\\
 \left[  Q_{i\alpha},Q_{j\beta}\right]  =\frac{i}%
{4}\left\{  \delta_{\alpha\beta}\varepsilon_{ijk}S_{k}+\delta_{ij}%
\varepsilon_{\alpha\beta\gamma}T_{\gamma}\right\} ,
\end{eqnarray}
where $\varepsilon_{ijk}$ is the completely antisymmetric tensor.
Clearly $\frak{su}\left(  2\right)  \times\frak{su}\left(  2\right)
$ is generated by the operators $\left\{
S_{i},T_{j}\right\}  $. The branching rule correspond to the representation
\begin{equation}
R=\left(  D_{1}\otimes D_{0}\right)  \oplus\left(  D_{0}\otimes
D_{1}\right)
\oplus\left(  D_{1}\otimes D_{1}\right)  ,\label{RD}%
\end{equation}
where $D_{1}$ denotes the adjoint representation of $\frak{su}(2)$
and $D_{0}$ the trivial representation. The two missing label
operators are found integrating the system
\begin{equation}
 \widehat{S}_{i}F   =\epsilon_{ijk}s_{k}\frac{\partial F}{\partial s_{j}%
}+\epsilon_{ijk}q_{kl}\frac{\partial F}{\partial q_{kl}}=0,\quad
\widehat{T}_{\alpha}F
=\epsilon_{\alpha\beta\gamma}t_{\gamma}\frac{\partial F}{\partial
t_{\beta}}+\epsilon_{\beta\gamma\mu}q_{\alpha\mu}\frac{\partial
F}{\partial q_{\beta\mu}}=0,\quad i=1,2,3\label{S2}%
\end{equation}
corresponding to the generators of the subalgebra. Five of the nine
independent solutions correspond to invariants of
$\frak{su}\left(
4\right)  $ and $\frak{su}\left(  2\right)  \times\frak{su}\left(  2\right)
$. The Casimir operators of $\frak{su}\left(  4\right)$ are
\[%
\begin{array}
[c]{ll}%
C_{2}= & \ s_{\alpha}s^{\alpha}+t_{\beta}t^{\beta}+4q_{\alpha\beta}%
q^{\alpha\beta},\\
C_{3}= & s_{\alpha}t_{\beta}q^{\alpha\beta}-4\varepsilon^{ijk}\varepsilon
^{\alpha\beta\gamma}q_{i\alpha}q_{j\beta}q_{k\gamma},\\
C_{4}= & 16\left\{  \varepsilon_{\alpha\beta\gamma}^{2}(q_{\alpha\beta}%
^{2}\left(  q_{\alpha\gamma}^{2}+q_{\gamma\beta}^{2}\right)  +2q_{\alpha
\alpha}^{2}\left(  q_{\alpha\gamma}^{2}+q_{\beta\alpha}^{2}\right)
-2q_{\alpha\alpha}q_{\alpha\beta}q_{\gamma\alpha}q_{\gamma\beta}+\left.
3q_{\alpha\beta}^{2}\left(  q_{\gamma\alpha}^{2}+q_{\gamma\gamma}^{2}\right)
\right)  \right.  \\
& \left.  +\sum_{a<\beta}\left(  3\left(  q_{\alpha\alpha}^{2}q_{\beta\beta
}^{2}+q_{\alpha\beta}^{2}q_{\beta\alpha}^{2}\right)  -2q_{\alpha\alpha
}q_{\beta\beta}q_{\alpha\beta}q_{\beta\alpha}\right)  \right\}  +\left(
s_{\alpha}s^{\alpha}\right)  ^{2}+\left(  t_{\beta}t^{\beta}\right)
^{2}+3s_{\alpha}s^{\alpha}t_{\beta}t^{\beta}\\
& +16q_{\alpha\beta}^{4}+2^{3}q_{\alpha\beta}^{2}\left(  s_{\alpha}s^{\alpha
}+t_{\beta}t^{\beta}\right)  +4\left\{  t_{\alpha}t_{\beta}q_{\gamma\alpha
}q_{\gamma\beta}+s_{\alpha}s_{\beta}q_{\alpha\gamma}q_{\beta\gamma
}-\varepsilon_{\alpha\beta\gamma}\varepsilon_{\mu\nu\rho}s_{\mu}t_{\alpha
}q_{\nu\beta}q_{\rho\gamma}\right\}  ,
\end{array}
\]

while $C_{21}=s_{\alpha}s^{\alpha},\quad C_{22}=t_{\beta}t^{\beta}$ are those
of the subalgebra. The contraction associated to the chain has the Levi
decomposition $\frak{g}=(\frak{su}(2)\times\frak{su}(2))\overrightarrow
{\oplus}_{D_{1}\otimes D_{1}}9L_{1}$, and it is not difficult to verify that
$\mathcal{N}\left(  \frak{g}\right)  =3$. Contraction of the invariants give
respectively%
\[%
\begin{array}
[c]{ll}%
C_{2}^{\prime}= & 4q_{\alpha\beta}q^{\alpha\beta},\\
C_{3}^{\prime}= & -4\varepsilon^{ijk}\varepsilon^{\alpha\beta\gamma}%
q_{i\alpha}q_{j\beta}q_{k\gamma},\\
C_{4}= & 16\left\{  \varepsilon_{\alpha\beta\gamma}^{2}(q_{\alpha\beta}%
^{2}\left(  q_{\alpha\gamma}^{2}+q_{\gamma\beta}^{2}\right)  +2q_{\alpha
\alpha}^{2}\left(  q_{\alpha\gamma}^{2}+q_{\beta\alpha}^{2}\right)
-2q_{\alpha\alpha}q_{\alpha\beta}q_{\gamma\alpha}q_{\gamma\beta}+\left.
3q_{\alpha\beta}^{2}\left(  q_{\gamma\alpha}^{2}+q_{\gamma\gamma}^{2}\right)
\right)  \right.  \\
& \left.  +\sum_{a<\beta}\left(  3\left(  q_{\alpha\alpha}^{2}q_{\beta\beta
}^{2}+q_{\alpha\beta}^{2}q_{\beta\alpha}^{2}\right)  -2q_{\alpha\alpha
}q_{\beta\beta}q_{\alpha\beta}q_{\beta\alpha}\right)  +16q_{\alpha\beta}%
^{4}\right\}  .
\end{array}
\]
To see that $\mathcal{F}=\left\{  C_{2},C_{3},C_{4},C_{21},C_{22},C_{3}%
^{\prime},C_{4}^{\prime}\right\}  $ is a functionally independent
set, we consider
the Jacobian with respect to the variable set $\left\{  s_{2},s_{3},t_{1}%
,t_{2},q_{11},q_{12},q_{23}\right\}  $ :
\begin{equation}
\frac{\partial(C_{21},C_{2},C_{3},C_{4},C_{2}^{\prime},C_{3}^{\prime}%
,C_{4}^{\prime})}{\partial(s_{2},s_{3},t_{1},t_{2},q_{11},q_{12},q_{23})}%
\neq0.
\end{equation}
Therefore the contraction method
provides two of the four available operators. To construct suitable
labelling operators, we take
the difference of the cubic invariants of $\frak{su}(4)$ and
$\frak{g}$. In this way we recover exactly the cubic operator $\Omega$ of
Moshinsky and Nagel \cite{Mo}:
\begin{equation}
C_{3}-C_{3}^{\prime}=\Omega=s_{\alpha}t_{\beta}q^{\alpha\beta}.
\end{equation}
This operator  is known to commute only with the fourth
order operator $\Phi$ defined by
\begin{equation}
\Phi=S_{i}S_{j}Q_{i\alpha}Q_{j\alpha}+Q_{i\alpha}Q_{i\beta}T_{\alpha}T_{\beta
}-\epsilon_{ijk}\epsilon_{\alpha\beta\gamma}S_{i}T_{\alpha}Q_{j\beta
}Q_{k\gamma}.\label{Op}%
\end{equation}
Algebraic manipulation of the fourth order operators of $\frak{su}(4)$ and
the contraction leads to the following result:
\begin{equation}
\Phi=\frac{1}{4}\left\{  C_{4}-C_{4}^{\prime}+C_{21}^{2}-C_{2}^{2}%
+C_{2}^{\prime2}-C_{21}\left(  C_{2}^{\prime}-C_{2}\right)
\right\}
.\label{Mos1}%
\end{equation}
This means that the commuting $\Omega-\Phi$ operators of
\cite{Mo} can be completely recovered by the contraction
associated to the embedding of spin-isospin subalgebra in
$\frak{su}\left(  4\right)  $. We observe that the remaining operators
are contained in the expression of $\Phi$, and cannot be isolated by
the contraction only.

\subsection{The $\frak{su}\left(  5\right)  \supset\frak{su}\left(  3\right)
\times\frak{su}\left(  2\right)  $ reduction}

Reduction chains of the type $\frak{su}\left(  p+q\right)  \supset
\frak{su}\left(  p\right)  \times\frak{su}\left(  q\right)  $ are very common
in particle physics, and constitute the natural generalization of the well
known breaking of symmetry of $\frak{su}\left(  3\right)  $ down to the
isospin and hypercharge $\frak{su}\left(  2\right)  \times\frak{u}\left(
1\right)  $. The unitary group $\frak{su}\left(  5\right)  $ is a central
object in the study how leptons get mass, and the adjoint representation of
$\frak{su}\left(  5\right)  $ with a vacuum value in the direction of the
$\frak{u}\left(  1\right)  $ generator is a good choice for the Higgs field
\cite{Geo} in the symmetry breaking $\frak{su}\left(  5\right)  \supset
\frak{su}\left(  3\right)  \times\frak{su}\left(  2\right)  \times
\frak{u}\left(  1\right)  $. Leaving aside $\frak{u}\left(  1\right)  $, the
corresponding reduction chain also presents some interest. In this case, the
number of needed labelling operators is
\[
n=\frac{1}{2}\left(  24-4-8-2-3-1\right)  =3.
\]
We will see that this chain can be solved using only the associated
contraction to the chain.\footnote{Whether this solution is optimal is another
question. Using the decompositions introduced later, the result can be
simplified to obtain more elementary labelling operators.}

\smallskip

In this work we will use the basis of $\frak{u}(N)$ given by the operators
$\left\{  E_{\mu\nu},F_{\mu\nu}\right\}  _{1\leq\mu,\nu\leq N}$ with the
constraints $E_{\mu\nu}+E_{\nu\mu}=0,\;F_{\mu\nu}-F_{\nu\mu}=0$. The
commutation relations over this basis are:
\begin{equation}
\begin{array}{ll}
\left[  E_{\mu\nu},E_{\lambda\sigma}\right]    & =E_{\nu\sigma}+E_{\lambda\nu
}-E_{\mu\sigma}-E_{\lambda\mu},\\
\left[  E_{\mu\nu},F_{\lambda\sigma}\right]    & =F_{\nu\sigma}+F_{\lambda\nu
}-F_{\mu\sigma}-F_{\lambda\mu},\\
\left[  F_{\mu\nu},F_{\lambda\sigma}\right]    & =E_{\nu\sigma}+E_{\mu\sigma
}-E_{\lambda\mu}-E_{\lambda\nu}.
\end{array}\label{b2}
\end{equation}
Since $\frak{u}\left(  N\right)  =\frak{su}\left(  N\right)  \oplus\mathbb{R}%
$, it follows at once that $\frak{u}\left(  N\right)  $ has $N$ independent
Casimir operators, one of them being central element, while the other $(N-1)$
invariants correspond to the simple part. To recover $\frak{su}\left(
N\right)  $, \  we take the Cartan subalgebra spanned by the vectors $H_{\mu
}=F_{\mu\mu}-F_{\mu+1,\mu+1}$ for $\mu=1..N-1$. The centre of $\frak{u}(N)$ is
then obviously generated by $\delta^{\mu\mu}F_{\mu\mu}$, and the remaining can
be deduced using an algebraic approach similar to the Gel'fand method. For the
case that interests us here, a maximal set of independent Casimir invariants
of $\frak{su}\left(  5\right)  $ is given by the coefficients $D_{k}$ of the
characteristic polynomial $\left|  iA_{5}-\lambda\mathrm{Id}_{5}\right|
=\lambda^{5}+\sum_{k=2}^{5}D_{k}\lambda^{5-k}$, where $A_{5}$ is the matrix
defined by
\begin{equation}
\left(
\begin{array}
[c]{ccccc}%
-iY_{1} & -e_{12}-i\,f_{12} & -e_{13}-i\,f_{13} & -e_{14}-i\,f_{14} &
-e_{15}-i\,f_{15}\\
e_{12}-i\,f_{12} & -iY_{2} & -e_{23}-i\,f_{23} & -e_{24}-i\,f_{24} &
-e_{25}-i\,f_{25}\\
e_{13}-i\,f_{13} & e_{23}-i\,f_{23} & -iY_{3} & -e_{34}-i\,f_{34} &
-e_{35}-i\,f_{35}\\
e_{14}-i\,f_{14} & e_{24}-i\,f_{24} & e_{34}-i\,f_{34} & -iY_{4} &
-e_{45}-i\,f_{45}\\
e_{15}-i\,f_{15} & e_{25}-i\,f_{25} & e_{35}-i\,f_{35} & e_{45}-i\,f_{45} &
-iY_{5}%
\end{array}
\right)  ,\label{Tv}%
\end{equation}
where the vectors $Y_{i}$ are given respectively by
\begin{equation}
\begin{array}
[c]{ll}%
Y_{1}=\frac{4}{5}h_{1}+\frac{3}{5}h_{2}+\frac{2}{5}h_{3}+\frac{1}{5}h_{4}, &
Y_{2}=-\frac{1}{5}h_{1}+\frac{3}{5}h_{2}+\frac{2}{5}h_{3}+\frac{1}{5}h_{4},\\
Y_{3}=-\frac{1}{5}h_{1}-\frac{2}{5}h_{2}+\frac{2}{5}h_{3}+\frac{1}{5}h_{4}, &
Y_{4}=-\frac{1}{5}h_{1}-\frac{2}{5}h_{2}-\frac{3}{5}h_{3}+\frac{1}{5}h_{4},\\
Y_{5}=-\frac{1}{5}h_{1}-\frac{2}{5}h_{2}-\frac{3}{5}h_{3}-\frac{4}{5}h_{4}. &
\end{array}
\end{equation}
Before symmetrization, the Casimir operators $D_{2},..,D_{5}$ obtained by
this method have $30,140,575$ and $1848$ terms. For this reduction chain, the
subalgebra $\frak{su}\left(  3\right)  \times\frak{su}\left(  2\right)  $ is
generated by $\left\{  \left\{  H_{1},H_{2},E_{\mu\nu},F_{\mu\nu}\right\}
_{1\leq\mu,\nu\leq3},\left\{  H_{4},E_{45},F_{45}\right\}  \right\}  $. In
particular, the inhomogeneous contraction $\frak{g}$ has Levi part isomorphic
to $\frak{su}\left(  3\right)  \times\frak{su}\left(  2\right)  $, and from
the properties of contractions it has at least four invariants. Since we need
three labelling operators, the conditions to solve the MLP by this method are
given. Contracting only the Casimir operators $D_{3},D_{4}$ and $D_{5}$, the
result leads to
\[%
\begin{array}
[c]{cc}%
D_{3}= & t^{4}D_{3}^{\prime}+\text{l.o.t.},\\
D_{4}= & t^{4}D_{4}^{\prime}+\text{l.o.t.},\\
D_{5}= & t^{4}D_{5}^{\prime}+\text{l.o.t.},
\end{array}
\]
where l.o.t. refers to thos terms having lower power in the contraction
variable $t$. The contracted invariants $D_{i}^{\prime}$ all have degree $i$
in the variables of the representation space complementary to the subalgebra
in $\frak{su}\left(  5\right)  $. We now define the operators
\begin{equation}
\begin{array}
[c]{cc}%
\Omega_{3}= & D_{3}-D_{3}^{\prime}-F_{3},\\
\Omega_{4}= & D_{4}-D_{4}^{\prime}-F_{4},\\
\Omega_{5}= & D_{5}-D_{5}^{\prime}-F_{5},
\end{array}
\end{equation}
where $F_{i}$ are the terms of the Casimir operators $D_{i}$ that only depend
on the variables of $\frak{su}\left(  3\right)  \times\frak{su}\left(
2\right)  $. The operator $\Omega_{3}$ has $89$ terms, $\Omega_{4}$ has $427$
and $\Omega_{5}$ $1618$ terms before symmetrization. A long computation shows
that $\Omega_{3},\Omega_{4}$ and $\Omega_{5}$ are independent on the $D_{i}$
and the Casimir operators of the subalgebra. To completely solve the labelling
problem, we still have to check the orthogonality conditions on the
symmetrized operators:\footnote{As usual, we denote the symmetrized and
non-symmetrized operators by the same symbol.}%
\begin{equation}
\begin{array}
[c]{ccc}%
\left[  \Omega_{3},\Omega_{4}\right]  =0, & \left[  \Omega_{3},\Omega
_{5}\right]  =0, & \left[  \Omega_{3},\Omega_{5}\right]  =0.
\end{array}
\end{equation}
This proves that $\left\{  \Omega_{3},\Omega_{4},\Omega_{5}\right\}  $ is a
complete set of labelling operators for the studied reduction chain. It should
be remarked that the main difficulty of this procedure is purely technical,
and corresponds to checking the commutator of the labelling operators. For
similar reduction chains of higher rank the procedure still remains valid.

\subsection{$\widehat{S}\left(  3\right)  \supset\frak{sl}\left(
2,\mathbb{R}\right)  \times\frak{so}\left(  2\right)  $}

First considered in \cite{Nie}, the invariance algebra of the
Schr\"{o}dinger equation in (N+1)-dimensional space time has attracted
considerable interest in recent physical literature (\cite{Bar2} and
references therein). The Schr\"{o}dinger algebra $\widehat{S}\left(  3\right)
$ in $\left(  3+1\right)  $-dimensional space-time is a $13$-dimensional Lie
algebra with non-trivial commutators%
\begin{equation}%
\begin{tabular}
[c]{ll}%
\multicolumn{2}{l}{$\left[  J_{\mu\nu},J_{\lambda\sigma}\right]  =\delta
_{\mu\lambda}J_{\nu\sigma}+\delta_{\nu\sigma}J_{\mu\lambda}-\delta_{\mu\sigma
}J_{\nu\lambda}-\delta_{\nu\lambda}J_{\mu\sigma},$}\\
$\left[  J_{\mu\nu},P_{\lambda}\right]  =\delta_{\mu\lambda}P_{\nu}%
-\delta_{\nu\lambda}P_{\mu},$ & $\left[  J_{\mu\nu},G_{\lambda}\right]
=\delta_{\mu\lambda}G_{\nu}-\delta_{\nu\lambda}G_{\mu},$\\
$\left[  P_{t},G_{\mu}\right]  =P_{\mu};$ & $\left[  K,P_{\mu}\right]
=-G_{\mu},$\\
$\left[  D,G_{\mu}\right]  =G_{\mu},$ & $\left[  D,P_{\mu}\right]  =-P_{\mu}%
,$\\
$\left[  D,K\right]  =2K,$ & $\left[  D,P_{t}\right]  =-2P_{t},$\\
$\left[  K,P_{t}\right]  =-D.$ & $\left[  P_{\mu},G_{\nu}\right]  =\delta
_{\mu\nu}M$%
\end{tabular}
\label{Sch}%
\end{equation}
over the basis $\left\{  J_{ij},P_{k},G_{k},K,D,P_{t},M\right\}  $, where
$J_{\mu\nu}+J_{\nu\mu}=0$ are rotations, $P_{\mu}$ are spatial translation
generators, $P_{t}$ the time translation, $G_{\mu}$ special Galilei
transformations, $D$ the generator of scale transformations, $K$ the generator
of galilean conformal transformations and $M$ commutes with all generators. It
follows from the brackets that the Levi decomposition is $\left(
\frak{so}(3)\oplus\frak{sl}\left(  2,\mathbb{R}\right)  \right)
\overrightarrow{\oplus}_{P}\frak{h}_{N}$, where the representation $P$ can be
identified with $(D_{\frac{1}{2}}\otimes\Lambda)\oplus D_{0}$, where
$D_{\frac{1}{2}}\otimes\Lambda$ is the tensor product of the standard
representations $D_{\frac{1}{2}}$ of $\frak{sl}(2,\mathbb{R})$ and $\Lambda$
of $\frak{so}(3)$, respectively, and denotes $D_{0}$ the trivial
representation. Let us consider the subalgebra $\frak{sl}\left(
2,\mathbb{R}\right)  \times\frak{so}\left(  2\right)  $ generated by $\left\{
J_{12},D,P_{t},M\right\}  $ and the corresponding reduction chain $\widehat
{S}\left(  3\right)  \supset\frak{sl}\left(  2,\mathbb{R}\right)
\times\frak{so}\left(  2\right)  $. In this case, the number of labelling
operators equals%
\[
n=\frac{1}{2}\left(  13-3-3-1-1-1\right)  +0=2.
\]
The system to be solved is therefore

\medskip%

\begin{align*}
\widehat{J}_{12}F  & =j_{2\sigma}\frac{\partial F}{\partial j_{1\sigma}%
}-j_{2\lambda}\frac{\partial F}{\partial j_{\lambda1}}-j_{1\sigma}%
\frac{\partial F}{\partial j_{2\sigma}}+j_{1\lambda}\frac{\partial F}{\partial
j_{\lambda2}}+r_{2}\frac{\partial F}{\partial r_{1}}-r_{1}\frac{\partial
F}{\partial r_{2}}+g_{2}\frac{\partial F}{\partial g_{1}}-g_{1}\frac{\partial
F}{\partial g_{2}}=0\nonumber\\
\widehat{D}F  & =2k\frac{\partial F}{\partial k}-2p_{t}\frac{\partial
F}{\partial p_{t}}+g_{\mu}\frac{\partial F}{\partial g_{\mu}}-r_{\mu}%
\frac{\partial F}{\partial r_{\mu}}=0\\
\widehat{K}F  & =-2k\frac{\partial F}{\partial d}-d\frac{\partial F}{\partial
p_{t}}-g_{\mu}\frac{\partial F}{\partial r_{\mu}}=0\\
\widehat{P}_{t}F  & =2p_{t}\frac{\partial F}{\partial d}+d\frac{\partial
F}{\partial k}+r_{\mu}\frac{\partial F}{\partial g_{\mu}}=0
\end{align*}
In \cite{C46} an algorithm was given to compute the Casimir
operators of the extended Schr\"odinger algebra. It is easy to see that $\widehat{S}\left(  3\right)  $ has
three invariants, one of them corresponding to the central charge $m$. The
other two can be chosen as%
\[%
\begin{array}
[c]{ll}%
C_{41}= & z^{2}j_{kl}^{2}+2z\left(  g_{k}p_{l}-g_{l}p_{k}\right)  j_{kl}%
+p_{k}^{2}g_{l}^{2}-2p_{k}p_{l}g_{k}g_{l},\\
C_{42}= & 2zkp_{k}^{2}+2ap_{t}g_{k}^{2}+z^{2}\left(  d^{2}-4kp_{t}\right)
-2g_{k}p_{k}dz+z^{2}j_{kl}^{2}+2z\left(  g_{k}p_{l}-g_{l}p_{k}\right)  j_{kl},
\end{array}
\]
where $1\leq k<l\leq3$. The invariants of $\frak{sl}\left(  2,\mathbb{R}%
\right)  \times\frak{so}\left(  2\right)  $ are simply $C_{21}=d^{2}-4kp_{t}$
and $C_{22}=j_{12}$. The contraction associated to the chain is obtained from
the scale transformations
\[
J_{kl}^{\prime}=\frac{1}{t}J_{kl}\;\left(  kl\neq12\right)  ,\;G_{i}^{\prime
}=\frac{1}{t}G_{i},\;P_{i}^{\prime}=\frac{1}{t}P_{i},\;M^{\prime}=\frac{1}%
{t}M.\;
\]
In this case, we still obtain a kind of inhomogeneous Lie algebra, the Levi
part of which is given by $\frak{sl}\left(  2,\mathbb{R}\right)  $. It
satisfies the preceding conditions, thus the MLP can be solved using the
invariants of the contraction. It follows at once that only $C_{41}$ and
$C_{42}$ are of interest, since the central invariant remains untouched by the
contraction. The corresponding contracted invariants are
\[%
\begin{array}
[c]{ll}%
C_{41}^{\prime}= & z^{2}j_{kl}^{2}+2z\left(  g_{k}p_{l}-g_{l}p_{k}\right)
j_{kl}+p_{k}^{2}g_{l}^{2}-2p_{k}p_{l}g_{k}g_{l},\;\\
C_{42}= & z^{2}j_{kl}^{2}+2z\left(  g_{k}p_{l}-g_{l}p_{k}\right)  j_{kl},
\end{array}
\]
where $j_{kl}\neq j_{12}$. We consider the differences%
\[%
\begin{array}
[c]{ll}%
O_{1}=C_{41}-C_{41}^{\prime}-m^{2}C_{22}^{2}= & j_{12}\left(  g_{1}p_{2}%
-p_{1}g_{2}\right)  z\\
O_{2}=C_{42}-C_{42}^{\prime}-m^{2}C_{21}^{2}-O_{1}= & 2z\left(  kp_{i}%
^{2}+p_{t}g_{i}^{2}-dg_{i}p_{i}\right)
\end{array}
\]
The independence of these operators from the Casimir operators is checked
by means of the Jacobian
\begin{equation}
\frac{\partial\left\{  C_{21},C_{22},C_{41},C_{42},M,O_{1},O_{2}\right\}
}{\partial\left\{  m,j_{12},d,p_{t},j_{23},g_{1},p_{2}\right\}  }\neq0.
\end{equation}
Finally, we compute the brackets%
\begin{equation}
\left[  O_{1},O_{2}\right]  =0,
\end{equation}
showing that the found operators commute.\newline The interest of this example is that the
Lie algebra used is not semisimple, showing that the method can be applied also to general Lie
algebras having non-trivial Levi decompositions\footnote{By this we mean that the maximal solvable ideal
is not reduced to zero.}. For higher dimensions of $n$ the procedure still works, although the refinement
developed later is probably more effective to find the suitable labelling operators.

\section{Chains solved only by contraction}

The contraction method constitutes a first approximation to
systematically solve the labelling problem in physical
applications of group theory. Even in cases with a high number of
labelling operators, this first step remains valid whenever the
conditions of theorem 1 are satisfied. In this sense, the method
can be applied for large classes of embeddings like
$\frak{sp}(2N)\supset\frak{sp}(2N-2)\times\frak{u}(1)$ or
$\frak{sp}(2N)\supset\frak{sp}(2N-2)\times\frak{su}(2)$, solutions
of which were developed in \cite{Bi}. Hovever, the contraction
fails if the contraction $\frak{g}$ has ``to few" invariants with
respect to the number of necessary labelling operators. This is
not an uncommon situation for reductive Lie algebras
$\frak{s}^{\prime}$ and semisimple $\frak{s}$ if the constraint
$\mathcal{N}(\frak{s})=\mathcal{N}(\frak{g})=n$ is given. These
reduction chains provide at most $n-1$ labelling operators. Up to
some special kinds of multiplets that can be solved using these
operators, for a general IR the remaining operator has to be
computed in some different manner. It is reasonable to think that
a refinement of the contraction procedure leads to the solution of
this obstruction, at least for a considerable number of embedding
chains. The reduction chain $G_{2}\supset
\frak{su}(2)\times\frak{su}(2)$ ($G_{2}$ being the rank two
exceptional Lie algebra) reflects the failure using only the
contracted invariants. In \cite{Hu1} a particular solution was
found by means of heavy algebra. It was moreover observed that
both labelling operators should have at least degree six. This
fact suggests to look more closely at the Casimir operator $C_{6}$
of $G_{2}$, in order to analyze whether some of the terms that
cancel during contraction provide an additional operator to that
of the contraction. This ansatz, corresponding to the second step
of the contraction method, is equal to study how the Casimir
operators decompose by contraction.

\section{Decomposing Casimir operators with respect to
contractions}

In this section we go further into a detailed analysis of the
decomposition procedure of Casimir operators. We point out that
the contraction associated to a reduction chain induces a
decomposition of the corresponding Casimir operators of
$\frak{s}$, which allow, among other properties, to determine the
invariants of the contraction $\frak{g}$ as the non-vanishing term
in the limit. However, other terms will also be relevant for the
missing label problem, and will provide additional labelling
operators.

In the general context developed earlier, let
$C_{p}(X_{1},...,X_{n})=\alpha^{i_{1}...i_{p}}X_{i_{1}}...X_{i_{p}}$
be a $p^{th}$-order Casimir operator of $\frak{s}$. Using a
contraction of the type (\ref{IWK}), the invariant over the
transformed basis takes the form
\begin{equation}
 F(\Phi_{t}(X_{1}),..,\Phi_{t}(X_{n})) =
t^{n_{i_{1}}+...+n_{i_{p}}}\alpha^{i_{1}...i_{p}}X_{i_{1}}...X_{i_{p}},
\end{equation}
where $n_{i_{j}}=0,1$. Taking the maximal power in $t$,
\begin{equation}
 M  = \max \left\{n_{i_{1}}+...+n_{i_{p}}\quad |\quad
\alpha^{i_{1}..i_{p}}\neq 0\right\},
\end{equation}
the limit
\begin{eqnarray*}
F^{\prime}(X_{1},..,X_{n}) =  \lim_{t\rightarrow \infty}
t^{-M}F(\Phi_{t}(X_{1}),...,\Phi_{t}(X_{n}))\\
 =  \sum_{n_{i_{1}}+...+n_{i_{p}}=M}
\alpha^{i_{1}...i_{p}}X_{i_{1}}...X_{i_{p}}
\end{eqnarray*}
provides a Casimir operator of degree $p$ of the contraction
$\frak{g}^{\prime}$, as previously used. Now, instead of
extracting only the term with the highest power of $t$, we
consider the whole decomposition
\begin{equation}
C_{p}=t^{M} C_{p}^{\prime}+ \sum_{\alpha}t^{\alpha} \Phi_{\alpha}
+  \Phi_{0}, \label{COZ}
\end{equation}
where $\alpha <M\leq p$ and $\Phi_{0}$ is a function of the
Casimir operators of the subalgebra $\frak{s}^{\prime}$ (these
generators have not been re-scaled). It is straightforward to
verify that $C_{p}^{\prime}$ is not only an invariant of the
contraction $\frak{g}$, but also a solution to the MLP. Equation
(\ref{COZ}) actually shows how a Casimir operator decomposes into
homogeneous polynomials in the variables of the subalgebra and the
complementary space over the original basis when a contraction is
performed. This first term corresponds to the first approximation
of \cite{C72}. The remaining terms are also individually of
interest to construct new labelling operators. Formally this fact
can be described as follows:

\begin{proposition}
The functions $\Phi_{\alpha}$ are solutions of the missing label
problem, that is, they satisfy the system
\begin{equation}
 \widehat{X}_{i}\Phi_{\alpha} = C_{ij}^{k}x_{k}\frac{\partial
\Phi_{\alpha}}{\partial x_{j}} = 0, \quad 1\leq i\leq s.
\label{ML1}
\end{equation}
\end{proposition}

\begin{proof}
Decomposition (\ref{COZ}) tells how a Casimir operator $C_{p}$ can
be rewritten as a sum of homogeneous polynomials $C_{p}^{\prime},
\Phi_{\alpha}$ with the property that $C_{p}^{\prime}$ is of
homogeneity degree $p-M$ in the variables
$\left\{x_{1},..,x_{s}\right\}$ associated to subalgebra
generators and degree $M$ in the remaining variables
$\left\{x_{s+1},..,x_{n}\right\}$ corresponding to the
representation space induced by the embedding. In similar way, any
$\Phi_{\alpha}$ is of degree $p-\alpha$ in the variables
$\left\{x_{1},..,x_{s}\right\}$ and $\alpha$ in the
$\left\{x_{s+1},..,x_{n}\right\}$. We denote this by simply saying
that these functions are of bi-degree $(p-\alpha,\alpha)$.

\smallskip

Now the equations (\ref{KS1}) corresponding to subalgebra
generators remain unaltered by the contraction procedure, since
the re-scaling of generators does not affect them. Thus for any
$1\leq i\leq s$ and any homogeneous polynomial $\Psi$ of bi-degree
$(p-q,q)$ we obtain
\begin{equation}
\widehat{X}_{i}\Psi=C_{ij}^{k}x_{k}\frac{\partial \Psi}{\partial
x_{j}}+C_{ij+s}^{k+s}x_{k+s}\frac{\partial \Psi}{\partial x_{j}},
\end{equation}
and the result is easily seen to be again a polynomial with the
same bi-degree. This means that evaluating $C_{p}=t^{M}
C_{p}^{\prime}+ \sum_{\alpha}t^{\alpha} \Phi_{\alpha} +  \Phi_{0}$
is a sum of polynomials of different bi-degree, and since $C_{p}$
is a Casimir operators, the only possibility is that each term is
a solution of the system. Therefore the functions $\Phi_{\alpha}$
are solutions of (\ref{KS1}).
\end{proof}

\medskip

The first question that arises from decomposition (\ref{COZ}) is
how many independent additional solutions we obtain. Since all
$\Phi_{\alpha}$ together sum the Casimir operator, some dependence
relations must exist.

\begin{lemma}
Let $C_{p}$ be a Casimir operator of $\frak{s}$ of order $p$.
Suppose that
\begin{equation}
C_{p}=\Phi_{(p-\alpha_{1},\alpha_{1})}+...+\Phi_{(p-\alpha_{q},\alpha_{q})},\quad
0\leq \alpha_{i}<\alpha_{i+1}\leq p
\end{equation}
is the decomposition of $C_{p}$ into homogeneous polynomials of
bi-degree $(p,q)$.
\begin{enumerate}

\item If $\Phi_{(0,p)}\neq 0$, then at most $q-2$
  polynomials $\Phi_{(p-\alpha_{j},\alpha_{j})}$ are functionally
independent on the Casimir operators of $\frak{s}$ and
$\frak{s}^{\prime}$.

\item If $\Phi_{(0,p)}=0$, then at most $q-1$
  polynomials $\Phi_{(p-\alpha_{j},\alpha_{j})}$ are functionally
independent on the Casimir operators of $\frak{s}$ and
$\frak{s}^{\prime}$.
\end{enumerate}
\end{lemma}

The proof is an immediate consequence of the fact  that
$\Phi_{(0,p)}$ is a function of the Casimir operators of the
subalgebra $\frak{s}^{\prime}$. The independence on the Casimir
operators of $\frak{s}^{\prime}$ does not imply in general that
the $\Phi_{(p-\alpha,\alpha)}$ obtained are all functionally
independent between themselves. The number of independent terms
depends on the representation $R$ induced by the reduction
\cite{C23}. In any case, at least one independent term is obtained
for any Casimir operator of degree at least three. In many cases,
however, we can take more terms of the same degree. This explains
why for certain reduction chains the labelling operators have the
same degree in the generators. If we find a functionally
independent set of solutions to system (\ref{KS1}), half the
labelling problem has been solved. In order to accomplish the
orthogonality requirements, we have to look for all commutators
among the symmetrized operators
$\Phi_{(p-\alpha_{j},\alpha_{j})}$. We denote by
$\Phi_{(p-\alpha_{j},\alpha_{j})}^{symm}$ the symmetrized
polynomial. Then
$\left[\Phi_{(p-\alpha_{j},\alpha_{j})}^{symm},\Phi_{(q-\alpha_{k},\alpha_{k})}^{symm}\right]$
 is a homogeneous polynomial of degree $p+q-1$, and also
constitutes a missing label operator. Actually this bracket is
expressible as sum of polynomials of different bi-degree, and
these terms constitute themselves labelling operators \cite{Jo}.

\medskip

The decomposition of Casimir operators leads to a first
generalization of the contraction method, resumed in the following
algorithmic procedure:

\begin{itemize}

\item Decompose the Casimir operators of $\frak{s}$ of degree
$p\geq 3$ with respect to the contraction determined by the
embedding.

\item Extract a maximal family of independent labelling operators.

\item Compute the commutators of all symmetrized polynomials $
\Phi_{(p-\alpha_{j},\alpha_{j})}^{symm}$ with $\alpha_{j}\neq 0$.

\item Extract $n$ operators that are functionally independent from
the Casimir operators of $\frak{s}$ and the subalgebra
$\frak{s}^{\prime}$ and commute among themselves.
\end{itemize}

The third step is reduced to pure computation. No simple method to
decide whether two missing label operators are mutually orthogonal
has been observed yet, although various symbolic routines have
been developed to compute these brackets (see e.g. \cite{DM}). For
certain special types of reduction chains it has been observed
that orthogonality follows at once from the second step. If no
solutions of bi-degree $(r,s)$ exists for some fixed $r+s=p+q$,
and if two labelling operators such that
$\left[\Phi_{(p-\alpha_{j},\alpha_{j})}^{symm},\Phi_{(q-\alpha_{k},\alpha_{k})}^{symm}\right]$
is a sum of polynomials of bi-degree $(s,r)$ are given, the
commutation is immediate. This idea was systematically developed
in \cite{Jo}. Observe that in the commutative frame, it would
suffice to show that no polynomial function of bi-degree $(r,s)$
is a solution to subsystem (\ref{KS1}). It should however the
remarked that the validity of this fact is reduced to quite
specific types of embeddings, and is therefore of no use in the
general labelling problem.

\section{Reduction chains solved with decomposition}

In this section we show how the decomposition of Casimir operators
of higher order provide solutions to missing label problem that
could not be solved completely by only using the contraction, or
for which no proposed set of labelling operators has been computed
yet. We insist on the fact that the main difficulty in the formal
approach to the MLP resides in obtaining a sufficient number of
(functionally) independent labelling operators, from which a
commuting set can be extracted.

\subsection{$G_{2}\supset \frak{su}(2)\times\frak{su}(2)$}

This chain was already observed in \cite{C72} to be unsolvable
using only the contraction invariants. In this case we need
$n=\frac{1}{2}\left(14-2-6-2\right)=2$ labelling operators, and
the inhomogeneous contraction $G_{2}\rightsquigarrow
(\frak{su}(2)\times\frak{su}(2))\overrightarrow{\oplus}_{R}8L_{1}$
preserves the number of invariants. The quadratic invariant being
discarded, the decomposition of the Casimir operator of degree six
must be used to obtain the pair of (commuting) labelling
operators. We consider the same tensor basis used in \cite{Hu1}
consisting of the generators
$\left\{j_{0},j_{\pm},k_{0},k_{\pm},R_{\mu,\nu}\right\}$ with
$\mu=\pm \frac{3}{2},\pm \frac{1}{2}$, $\nu=\pm\frac{1}{2}$. The
generators $R_{\mu,\nu}$ are related to an irreducible tensor
representation $R$ of $\frak{su}(2)\times\frak{su}(2)$ of order
eight. In this case, the contraction $G_{2}\rightsquigarrow
(\frak{su}(2)\times\frak{su}(2))\overrightarrow{\oplus}_{R}8L_{1}$
is obtained considering the transformations:
\[
j_{0}^{\prime }=j_{0},\;j_{\pm }^{\prime }=j_{\pm
},\;k_{0}^{\prime
}=k_{0},k_{\pm }^{\prime }=k_{\pm },\;R_{\mu ,\nu }^{\prime }=\frac{1}{t}%
R_{\mu ,\nu }.
\]
Decomposing now the Casimir operators $C_{2}$ and $C_{6}$ over the
transformed basis, we get the following
\begin{equation}
\begin{array}[l]{ll}
C_{2}=t^{2}C_{(2,0)}+ C_{(0,2)}, & \\
C_{6}=t^{6}C_{(6,0)}+t^{4}C_{(4,2)}+t^{2}C_{(2,4)}+C_{(0,6)},&
\end{array}
\end{equation}
where  $C_{(0,2)},C_{(0,6)}$  are functions of the Casimir
operators of $\frak{su}(2)\times\frak{su}(2)$. Now it can be
verified that
\begin{equation}
\frac{\partial{\left(C_{2},C_{6},C_{21},C_{22},C_{(2,4)},C_{(4,2)}\right)}}
{\partial{\left(k_{0},k_{-},j_{0},j_{+},R_{\frac{3}{2},\frac{1}{2}},R_{-\frac{3}{2},\frac{1}{2}}\right)}}\neq
0,
\end{equation}
where $C_{21}$ and $C_{22}$ are the quadratic Casimir operators of
$\frak{su}(2)\times\frak{su}(2)$. This provides us with six
independent operators. A long and tedious computation, due to the
quite high number of terms before and after symmetrization, shows
moreover that the chosen operators commute:
\begin{equation}
\begin{array}[l]{ll}
\left[C_{i},C_{(2,4)}\right]=\left[C_{i},C_{(4,2)}\right]=0, &
i=2,6\\
\left[C_{(4,2)},C_{(2,4)}\right]=0.&
\end{array}
\end{equation}
Therefore the set
$\left\{C_{2},C_{6},C_{21},C_{22},C_{(2,4)},C_{(4,2)}\right\}$ can
be taken to solve the labelling problem.

\smallskip

It should be remarked that a direct comparison with the operators
obtained in \cite{Hu1} is quite difficult, for various reasons. At
first, there the scalars in the enveloping algebra were
considered, not symmetrizations of functions, which implies that
lower order terms where considered when explicitly indicating the
labelling operators. On the other hand, we have only distinguished
the bi-degree, that is, the degree of the polynomials in the
variables of the $\frak{su}(2)\times\frak{su}(2)$ subalgebra and
the tensor representation $R$, while in \cite{Hu1} the order with
respect to any of the copies of $\frak{su}(2)$ was considered,
resulting in operators labelled with three indices. Therefore the
operators $C_{(p,q})$ considered here correspond to the sum of
several scalars there. In addition, our solution contains the term
$C^{(114)}$ excluded in \cite{Hu1},\footnote{This is a scalar
having degree one in each of the copies of $\frak{su}(2)$ and four
in the $R_{\mu,\nu}$ generators.} confirming that the pair of
commuting operators obtained above is different from that found
previously. We also remark that a further distinction of the
degrees of the polynomials $\Phi_{(a,b)}^{symm}$ in the variables
of the $\frak{su}(2)$ copies is not possible due to the
contraction.

\subsection{The chain $\frak{so}\left(  7\right)  \supset
\frak{so}\left(  5\right)  \times\frak{so}\left(  2\right)  $}

Reduction chains of orthogonal algebras have been analyzed in
\cite{Bi} from the algebraic point of view, proving interesting formulae based
on symmetric and antisymmetric tensor operators. We show that the
decomposition of the Casimir operators is also a valid approach to the
problem. For the reduction chain $\frak{so}\left(  7\right)  \supset
\frak{so}\left(  5\right)  \times\frak{so}\left(  2\right)  $ the number of
missing label operators is
\[
n=\frac{1}{2}\left(  21-3-10-2-1-1\right)  +0=2.
\]
As follows from the work of Gel'fand, the Casimir operators of orthogonal Lie
algebras can be recovered in a quite simple manner using the generic matrix of
standard representation and evaluating the corresponding characteristic
polynomial. The operators are then recovered by the symmetrization procedure.
Taking the usual basis generated by the $\frac{7}{2}(7-1)$ operators
$E_{\mu\nu}=-E_{\nu\mu}$ with brackets:
\begin{equation}
\left[  E_{\mu\nu},E_{\lambda\sigma}\right]  =E_{\nu\sigma}+E_{\lambda\nu
}-E_{\mu\sigma}-E_{\lambda\mu},
\end{equation}
the Casimir operators are obtained using the formula \cite{Ge}:
\begin{equation}
P(\lambda)=\left|  M-\lambda\mathrm{Id}_{7}\right|  =\lambda^{7}+C_{2}%
\lambda^{5}+C_{4}\lambda^{3}+C_{6}\lambda,
\end{equation}
$M$ being the matrix
\begin{equation}
M=\left(
\begin{array}
[c]{ccccccc}%
0 & -e_{12} & -e_{13} & -e_{14} & -e_{15} & -e_{16} & -e_{17}\\
e_{12} & 0 & -e_{23} & -e_{24} & -e_{25} & -e_{26} & -e_{27}\\
e_{13} & e_{23} & 0 & -e_{34} & -e_{35} & -e_{36} & -e_{37}\\
e_{14} & e_{24} & e_{34} & 0 & -e_{45} & -e_{46} & -e_{47}\\
e_{15} & e_{25} & e_{35} & e_{45} & 0 & -e_{56} & -e_{57}\\
e_{16} & e_{26} & e_{36} & e_{46} & e_{56} & 0 & -e_{67}\\
e_{17} & e_{27} & e_{37} & e_{47} & e_{57} & e_{67} & 0
\end{array}
\right)  .\label{ST}%
\end{equation}
For the chain considered, the $\frak{so}\left(  5\right)  \times
\frak{so}\left(  2\right)  $ subalgebra is generated by the operators $E_{ij}$
with $1\leq i,j\leq5$ and $E_{67}$. Therefore the contraction related to the
MLP is determined by the transformations%
\[
E_{ij}^{\prime}=\frac{1}{t}E_{ij},\;E_{67}^{\prime}=\frac{1}{t}E_{67},\;1\leq
i,j\leq5.
\]
As usual, the quadratic Casimir operator is of no use, we therefore decompose
the remaining ones:%
\[%
\begin{array}
[c]{cc}%
C_{4}= & C_{\left[  4,0\right]  }+{t^{2}}C_{\left[  2,2\right]
}+{t^{4}}C_{\left[  0,4\right]  },\\
C_{6}= & C_{\left[  6,0\right]  }+{t^{2}}C_{\left[  4,2\right]
}+{t^{4}}C_{\left[  2,4\right]  }.
\end{array}
\]
The functions $C_{\left[  4,0\right]  }$ and $C_{\left[  6,0\right]  }$ are
functions of the subalgebra generators, and therefore not further interesting.
The remaining operators $C_{\left[  2,2\right]  },C_{\left[  0,4\right]
},C_{\left[  4,2\right]  }$ and $C_{\left[  2,4\right]  }$ have $140,30,420$
and $390$ terms before symmetrization. We observe that each Casimir operator
provides at most one independent labelling operator, which can be taken either
as  $C_{\left[  2,2\right]  }$ or $C_{\left[  0,4\right]  }$ for $C_{4}$ and
$C_{\left[  4,2\right]  }$ or $C_{\left[  2,4\right]  }$ for $C_{6}$. Taking
for example the pair $C_{\left[  2,2\right]  }$ and $C_{\left[  2,4\right]  }%
$, we check their independence on the invariants of $\frak{so}\left(
7\right)  $ and the subalgebra:%
\[
\frac{\partial\left\{  C_{21},C_{41},C_{22},C_{2},C_{4},C_{6},C_{\left[
2,2\right]  },C_{\left[  2,4\right]  }\right\}  }{\partial\left\{
e_{12},e_{13},e_{16},e_{34}.e_{46},e_{56},e_{57,}e_{67}\right\}  }\neq0,
\]
where $C_{21}$ and $C_{41}$ are the Casimir operators of $\frak{so}\left(
5\right)  $ and $C_{22}=e_{67}$. A routine computation shows that the
orthogonality constraints are verified:%
\begin{equation}
\begin{array}
[c]{lll}%
\left[  C_{\left[  2,2\right]  },C_{21}\right]  =0, & \left[  C_{\left[
2,2\right]  },C_{41}\right]  =0, & \left[  C_{\left[  2,2\right]  }%
,C_{22}\right]  =0,\\
\left[  C_{\left[  2,4\right]  },C_{21}\right]  =0, & \left[  C_{\left[
2,4\right]  },C_{41}\right]  =0, & \left[  C_{\left[  2,4\right]  }%
,C_{22}\right]  =0,\\
\left[  C_{\left[  2,2\right]  },C_{p}\right]  =0, & \left[  C_{\left[
2,4\right]  },C_{p}\right]  =0, & \left[  C_{\left[  2,2\right]  },C_{\left[
2,4\right]  }\right]  =0,
\end{array}
\end{equation}
for $p=2,4,6$. The last brackets shows that the labelling operators commute,
therefore they constitute an admissible solution to the MLP given by the
embedding.

\subsection{The chain $\frak{sp}(6)>\frak{su}(3)\times
\frak{u}(1)$}

The unitary reduction of the symplectic Lie algebra of rank three
has found ample applications in the nuclear collective model
\cite{RoG}. In this case, nuclear states are classified by means
of irreducible representations of $\frak{sp}(6)$ reduced with
respect to the unitary subalgebra $\frak{su}(3)\times
\frak{u}(1)$. Since the induced representations are not
multiplicity free, we have to add $n=3$ labelling operators to
distinguish the states. Generating functions for this chain were
studied in \cite{GaK}, but without obtaining explicitly the three
required operators. In this section, we will determine a commuting
set of labelling operators that solves the MLP for this reduction.
As we shall see, this case cannot be solved using only the
invariants of the associated contraction.

\smallskip

We will use the Racah realization for the symplectic Lie algebra
$\frak{sp}\left(6,\mathbb{R}\right)$. We consider the generators
$X_{i,j}$ with $-3\leq i,j\leq 3$ satisfying the condition
\begin{equation}
X_{i,j}+\varepsilon_{i}\varepsilon_{j}X_{-j,-i}=0,
\end{equation}
where $\varepsilon_{i}={\rm sgn}\left( i\right) $. Over this
basis, the brackets are given by
\begin{equation}
\left[  X_{i,j},X_{k,l}\right]  =\delta_{jk}X_{il}-\delta_{il}X_{kj}%
+\varepsilon_{i}\varepsilon_{j}\delta_{j,-l}X_{k,-i}-\varepsilon
_{i}\varepsilon_{j}\delta_{i,-k}X_{-j,l}, \label{Kl3}
\end{equation}
where $-3\leq i,j,k,l\leq 3$. The three Casimir operators
$C_{2},C_{4},C_{6}$ of $\frak{sp}\left( 6,\mathbb{R}\right)  $ are
easily obtained as the coefficients of the characteristic
polynomial
\begin{equation}
\left |  A-T {\rm Id}_{6}\right| =T^{6}+
C_{2}T^{4}+C_{4}T^{2}+C_{6},\label{LOL}
\end{equation}
where
\begin{equation}
A=\left(
\begin{array}
[c]{cccccc}%
x_{1,1} & x_{2,1} & x_{3,1} & -I x_{-1,1} & -I x_{-1,2} & - I x_{-1,3}\\
x_{1,2} & x_{2,2} & x_{3,2} & -I x_{-1,2} & -I x_{-2,2} & -I x_{-2,3}\\
x_{1,3} & x_{2,3} & x_{3,3} & -I x_{-1,3} & -I x_{-2,3} & -I
x_{-3,3}\\
I x_{1,-1} & I x_{1,-2} & I x_{1,-3} & -x_{1,1} & -x_{1,2} & -x_{1,3}\\
I x_{1,-2} & I x_{2,-2} & I x_{2,-3} & -x_{2,1} & -x_{2,2}
&-x_{2,3}\\
I x_{1,-3} & I x_{2,-3} & I x_{3,-3} & -x_{3,1} & -x_{2,3}
&-x_{3,3}
\end{array}
\right). \label{M0}
\end{equation}
The symmetrized operators give the usual polynomials in the
enveloping algebra. Since the unitary algebra $\frak{u}(3)$ is
generated by $\left\{X_{i,j} | 1\leq i,j\leq 3\right\}$, in order
to write $\frak{sp}\left( 6,\mathbb{R}\right)$ in a
$\frak{su}(3)\times \frak{u}(1)$ basis, it suffices to replace the
diagonal operators $X_{i,i}$ by suitable linear combinations.
Taking $H_{1}=X_{1,1}-X_{2,2},\; H_{2}=X_{2,2}-X_{3,3}$ and
$H_{3}=X_{1,1}+X_{2,2}+X_{3,3}$ we obtain the Cartan subalgebra of
$\frak{su}(3)$, while $H_{3}$ commutes with all $X_{i,j}$ with
positive indices $i,j$. The invariants over this new basis are
simply obtained replacing the $x_{i,i}$ by the corresponding
linear combinations of $h_{i}$. The contraction $\frak{sp}(6)
\rightsquigarrow (\frak{su}(3)\times
\frak{u}(1))\overrightarrow{\oplus}_{R}12L_{1}$, where $R$ is the
complementary to $({\rm ad}(\frak{su}(3)\otimes (1))$ in the
adjoint representation of $\frak{sp}(6)$:\footnote{More precisely,
$R$ decomposes into a sextet and antisextet with $\frak{u}(1)$
weight $\pm 1$ and a singlet with $\frak{u}(1)$ weight $1$.}
\[
{\rm ad}\frak{sp}(6)= ({\rm ad}\frak{su}(3)\otimes (1))\oplus R.
\]
The contraction is determined by the transformations
\begin{equation}
H_{i}^{\prime}=H_{i},\; X_{i,j}^{\prime}=X_{i,j},\;
X_{-i,j}^{\prime}=\frac{1}{t}X_{-i,j},\;
X_{i,-j}^{\prime}=\frac{1}{t}X_{i,-j},\quad 1\leq i,j\leq 3.
\end{equation}
The contraction $(\frak{su}(3)\times
\frak{u}(1))\overrightarrow{\oplus}_{R}12L_{1}$ satisfies
$\mathcal{N}=3$, thus has 3 Casimir operators that can be obtained
as contraction of $C_{2},C_{4},C_{6}$. Note however that $n=3$,
thus the invariants of the contraction will provide at most two
independent missing label operators. This means that using only
the contraction, we cannot solve the MLP for this chain. In order
to find a third labelling operator, we have to consider the
decomposition of the fourth and sixth order Casimir operators of
$\frak{sp}(6)$. Over the preceding transformed basis we obtain:
\begin{equation}
\begin{array}[l]{ll}
C_{4}= t^{4} C_{(4,0)}+ t^{2} C_{(2,2)}+ C_{(0,4)},& \\
C_{6}= t^{6} C_{(6,0)}+ t^{4} C_{(4,2})+ t^{2} C_{(2,4)}+
C_{(0,6)},&
\end{array}
\end{equation}
where $C_{(k,l)}$ denotes a homogeneous polynomial of $k$ in the
variables of $R$ and degree $l$ in the variables of the unitary
subalgebra. The $C_{(0,k)}$ are functions of the Casimir operators
of $\frak{su}(3)\times \frak{u}(1)$, and therefore provide no
labelling operators. We remark that, before symmetrization,
$C_{(2,2)}$ has $126$ terms, $C_{(2,4)}$ $686$ terms, and
$C_{(4,2)}$ 444 terms. The symmetrized operators
$C_{(2,2)},C_{(4,2)}$ and $C_{(2,4)}$ can be added to the Casimir
operators of $\frak{sp}(6)$ and the subalgebra $\frak{su}(3)\times
\frak{u}(1)$, and the 9 operators can be seen to be functionally
independent.
\begin{equation}
\begin{array}
[c]{llll}%
\left[  C_{i},C_{(2,2)}\right]  =0, & \left[  C_{i},C_{(4,2)}\right]  =0, &
\left[  C_{i},C_{(2,4)}\right]  =0, & i=2,4,6\\
\left[  C_{(2,2)},C_{(4,2)}\right]  =0, & \left[  C_{(2,2)},C_{(2,4)}\right]
=0, & \left[  C_{(2,4)},C_{(4,2)}\right]  =0. &
\end{array}
\end{equation}

For symplectic algebras of higher rank, the decomposition method still provides the
required labelling operators. As expected, the main difficulty lies in the computation
of the brackets of the operators, where the number of terms increases exponentially.

\section{Complete solutions}

Even if the decomposition of the Casimir operators constitute a
great improvement of the contraction method in the MLP, there
exist reduction chains where the problem cannot be solved
completely. In this case, the failure is related to the additional
orthogonality constraint. Even if in almost any case we can find a
sufficient number of independent labelling operators, these do not
provide linear combinations that lead to mutually orthogonal
operators. As already observed, we cannot introduce a further
refinement of the decomposition without altering the contraction,
and, therefore the reduction chain itself. In order to obtain
integrity bases, we must look for new labelling operators that do
not arise from the decomposition. In these cases, we are forced to find additional
solutions to the subsystem of PDEs associated to the subalgebra generators.
The search for such operators can be simplified if we require them to have
a specific bi-degree. This requirement has been systematically used in
the literature.

\subsection{The nuclear surfon model}

The reduction chain $\frak{so}(5)\supset \frak{so}(3)$ appears in many
applications involving the subalgebra of angular momentum, and also
plays an important role in the Interacting Boson Model \cite{Ia,Oss}, where it
appears in the chains $\frak{u}(5)\supset\frak{so}(5)\supset\frak{so}(3)\supset\frak{so}(2)$
and $\frak{so}(6)\supset\frak{so}(5)\supset\frak{so}(3)\supset\frak{so}(2)$. The
corresponding $n=2$ missing label problem has been
analyzed in \cite{Me}, where two commuting missing label operators
of degrees four and six were found. The constructed two operators, using
heavy algebraic methods, are of lowest possible degree to solve the labelling problem.
The general shape of the labelling operators was however not found. Combining the
decomposition of Casimir operators with the traditional analytical approach to the MLP,
we give the complete solution for this embedding. In particular, it is verified that the
pair of commuting operators found in \cite{Me} correspond to the simplest possible
solution, and the conjecture on the degree of these operators is confirmed.

\medskip

We choose the basis of the orthogonal Lie algebra $\frak{so}(5)$ to consist of
generators $\left\{L_{0},L_{1},L_{-1}\right\}$ with brackets
$\left[L_{0},L_{\pm 1}\right]=\pm L_{\pm 1},\;
\left[L_{1},L_{-1}\right]=2L_{0}$ together with an irreducible
tensor representation $Q_{\mu}\; (\mu=-3..3)$ of dimension seven. The brackets of
$\frak{so}(5)$ over this basis are specified given in Table 3.

\begin{table}
\caption{\label{Tab1} $\frak{so}(5)$ brackets in a
$\frak{so}(3)$ basis.}
\begin{tabular}[c]{@{}c|cccccccccc}%
$\left[  {}\right]  $ & $Q_{3}$ & $Q_{2}$ & $Q_{1}$ & $Q_{0}$ &
$Q_{-1}$ & $Q_{-2}$ & $Q_{-3}$\\\hline
 $L_{0}$ & $3Q_{3}$ &
$2Q_{2}$ & $Q_{1}$ & $0$ & $-Q_{-1}$ & $-2Q_{-2}$ &
$-3Q_{-3}$\\
$L_{1}$ & $0$ & $6Q_{3}$ & $Q_{2}$ & $2Q_{1}$ & $6Q_{0}$ &
$10Q_{-1}$ &
$Q_{-2}$\\
$L_{-1}$ & $Q_{2}$ & $10Q_{1}$ & $6Q_{0}$ & $2Q_{-1}$ & $Q_{-2}$ &
$6Q_{-3}$ &
$0$\\
$Q_{3}$ & $0$ & $0$ & $0$ & $Q_{3}$ & $Q_{2}$ & $10Q_{1}+15L_{1}$
&
$5Q_{0}-15L_{0}$\\
$Q_{2}$ &  & $0$ & $-6Q_{3}$ & $-Q_{2}$ & $-15L_{1}$ &
$30Q_{0}+60L_{0}$ &
$10Q_{-1}-15L_{-1}$\\
$Q_{1}$ &  &  & $0$ & $3L_{1}-Q_{1}$ & $-3L_{0}-3Q_{0}$ &
$15L_{-1}$ &
$Q_{-2}$\\
$Q_{0}$ &  &  &  & $0$ & $-Q_{-1}-3L_{-1}$ & $-Q_{-2}$ & $Q_{-3}$\\
$Q_{-1}$ &  &  &  &  & $0$ & $-6Q_{-3}$ & $0$\\
$Q_{-2}$ &  &  &  &  &  & $0$ & $0$ \\\hline
\end{tabular}
\end{table}

According to the computations developed in \cite{Me}, the Casimir operators of $\frak{so}(3)$ and
$\frak{so}(5)$ are given respectively by the following (unsymmetrized) polynomials:

\[%
\begin{array}
[c]{ll}%
C_{21}= & l_{0}^{2}+l_{1}l_{-1},\\
C_{2}= & l_{0}^{2}+l_{1}l_{-1}-\frac{2}{5}\left(  q_{3}q_{-3}+q_{1}%
q_{-1}\right)  +\frac{1}{15}q_{2}q_{2}+q_{0}^{2},\\
C_{4}= & \frac{1}{6}\left(  l_{-1}q_{1}-l_{1}q_{-1}+\frac{l_{1}l_{-1}}%
{2}\right)  q_{0}^{2}+\frac{1}{6}\left(  q_{3}q_{-1}q_{-2}+q_{2}q_{1}%
q_{-3}+\frac{1}{3}\left(  q_{1}^{2}q_{-2}+q_{2}q_{-1}^{2}\right)  \right)
q_{0}\\
& -\left(  \frac{1}{9}l_{-1}q_{-1}+\frac{1}{6}l_{0}q_{-2}+\frac{2}{9}%
l_{1}q_{-3}\right)  q_{1}^{2}+\left(  \frac{2}{9}l_{-1}q_{3}+\frac{1}{9}%
l_{1}q_{1}-\frac{1}{6}l_{0}q_{2}\right)  q_{-1}^{2}\\
& \frac{1}{3}\left(  \frac{1}{20}q_{2}q_{-2}-q_{1}q_{-1}-3l_{-1}q_{1}+\frac
{7}{4}q_{0}^{2}+3l_{1}q_{-1}+\frac{1}{5}q_{3}q_{-3}\right)  l_{0}^{2}+\frac
{1}{18}l_{0}q_{3}q_{-1}q_{-2}\\
& +\frac{1}{18}l_{0}q_{2}q_{1}q_{-3}+\frac{3}{100}q_{3}^{2}q_{-3}^{2}+\frac
{1}{12}\left(  q_{-1}^{2}-3l_{-1}q_{-1}+3l_{0}q_{-2}+q_{1}q_{-3}-q_{0}%
q_{-2}\right)  l_{1}^{2}\\
& +\frac{1}{4}\left(  l_{1}q_{2}q_{-3}-\frac{1}{9}l_{-1}q_{2}q_{-1}+\frac
{1}{9}l_{1}q_{1}q_{-2}-l_{-1}q_{3}q_{-2}\right)  q_{0}+\frac{q_{2}^{2}%
q_{-2}^{2}}{675}+l_{0}^{3}q_{0}-l_{0}q_{0}^{3}\\
& +\frac{1}{12}\left(  \left(  q_{2}q_{-3}-q_{1}q_{-2}\right)  l_{1}%
l_{0}+\left(  -q_{3}q_{-2}+q_{2}q_{-1}\right)  l_{-1}l_{0}\right)  -\frac
{1}{6}\left(  l_{-1}q_{1}-l_{1}q_{-1}\right)  l_{0}q_{0}\\
& +\frac{1}{5}\left(  \frac{7}{6}q_{1}q_{-1}+\frac{1}{20}q_{2}q_{-2}\right)
q_{-3}q_{3}+\frac{1}{12}\left(  3l_{0}q_{2}+q_{3}q_{-1}+q_{1}^{2}+3l_{1}%
q_{1}-q_{2}q_{0}\right)  l_{-1}^{2}\\
& +\left(  \frac{l_{-1}q_{1}}{3}-\frac{l_{1}q_{-1}}{3}\right)  q_{-3}%
q_{3}+\frac{1}{6}\left(  \frac{1}{10}l_{1}l_{-1}-\frac{q_{-2}q_{2}}{6}%
l_{1}q_{-1}+\frac{2}{3}l_{0}q_{0}+\frac{1}{6}l_{-1}q_{1}\right)  q_{3}q_{-3}\\
& -\frac{1}{36}\left(  q_{2}^{2}q_{-1}q_{-3}-l_{1}q_{3}q_{-2}^{2}+q_{3}%
q_{1}q_{-2}^{2}+l_{-1}q_{2}^{2}q_{-3}\right)  -\frac{1}{9}\left(  q_{1}%
^{3}q_{-3}+q_{3}q_{-1}^{3}\right)  \\
& -\frac{5}{108}q_{1}^{2}q_{-1}^{2}-\frac{1}{12}\left(  l_{1}l_{-1}-\frac
{34}{3}l_{0}q_{0}\right)  q_{-1}q_{1}-\frac{q_{-2}q_{2}}{540}\left(
q_{1}q_{-1}+36q_{0}^{2}\right)  +\frac{1}{4}l_{1}^{3}q_{-3}\\
& -\left(  \frac{11}{60}l_{1}l_{-1}+\frac{1}{2}l_{0}q_{0}\right)  q_{-3}%
q_{3}-\frac{9}{6}l_{1}l_{-1}l_{0}q_{0}-\frac{3}{5}q_{0}^{2}q_{3}q_{-3}%
-\frac{1}{4}l_{-1}^{3}q_{3}\\
& \\
& \\
&
\end{array}
\]

For this Lie algebras, the transformations (\ref{TB}) defining the associated
contraction $\frak{g}$ are given by $L_{i}^{\prime}=L_{i},\;
Q_{\mu}^{\prime}=\frac{1}{t}Q_{\mu}$. The inhomogeneous contraction has an
Abelian radical of dimension seven, which implies that the
invariants will only depend on the $q_{\mu}$-variables \cite{C23}.
It is easy to verify that $\mathcal{N}(\frak{g})=4$,
and from the four Casimir operators, two can be obtained by
contraction of the invariants $C_{2}$ and $C_{4}$ of $\frak{so}(5)$.
We decompose $C_{4}$ with respect to the given contraction, and obtain
\begin{equation}
C_{4}= \left[4,0\right]+\left[3,1\right]+\left[2,2\right]+\left[1,3\right],
\end{equation}
where the operators $\left[i,j\right]$ are defined as follows:
\[%
\begin{array}
[c]{ll}%
\left[  1,3\right]  = & \frac{1}{4}l_{0}l_{1}^{2}q_{-2}-\frac{3}{2}l_{0}%
l_{1}l_{-1}q_{0}+\frac{1}{4}l_{1}^{3}q_{-3}-l_{0}^{2}l_{-1}q_{1}+l_{0}%
^{2}l_{1}q_{-1}+l_{0}^{3}q_{0}+\frac{1}{4}l_{0}l_{-1}^{2}q_{2}-\frac{1}%
{4}l_{-1}^{3}q_{3}\\
& -\frac{1}{4}l_{1}^{2}l_{-1}q_{-1}+\frac{1}{4}l_{1}l_{-1}^{2}q_{1}\\
& \\
\left[  2,2\right]  = & -\frac{1}{12}l_{-1}^{2}q_{2}q_{0}+\frac{1}{12}%
l_{0}l_{-1}q_{2}q_{-1}+\frac{1}{6}l_{0}l_{1}q_{0}q_{-1}+\frac{1}{12}l_{1}%
^{2}q_{1}q_{-3}-\frac{1}{3}l_{0}^{2}q_{1}q_{-1}+\frac{7}{12}l_{0}{}^{2}%
q_{0}^{2}\\
& +\frac{1}{60}l_{1}l_{-1}q_{2}q_{-2}+\frac{1}{12}l_{-1}^{2}q_{3}q_{-1}%
-\frac{1}{12}l_{0}l_{-1}q_{3}q_{-2}+\frac{1}{12}l_{1}l_{-1}q_{0}^{2}-\frac
{1}{12}l_{1}^{2}q_{0}q_{-2}+\frac{1}{12}l_{1}^{2}q_{-1}^{2}\\
& -\frac{1}{6}l_{0}l_{-1}q_{1}q_{0}-\frac{1}{12}l_{1}l_{-1}q_{1}q_{-1}%
+\frac{1}{12}l_{-1}^{2}q_{1}^{2}+\frac{1}{15}l_{0}^{2}q_{3}q_{-3}+\frac{1}%
{60}l_{0}^{2}q_{2}q_{-2}-\frac{1}{12}l_{0}l_{1}q_{1}q_{-2}\\
& -\frac{11}{60}l_{1}l_{-1}q_{3}q_{-3}+\frac{1}{12}l_{0}l_{1}q_{2}q_{-3}
\end{array}
\]

\[%
\begin{array}
[c]{ll}
\left[  3,1\right]  = & \frac{1}{4}l_{1}q_{2}q_{0}q_{-3}+\frac{1}{9}l_{0}%
q_{2}q_{0}q_{-2}-l_{0}q_{0}^{3}-\frac{1}{2}l_{0}q_{3}q_{0}q_{-3}+\frac{17}%
{18}l_{0}q_{1}q_{0}q_{-1}+\frac{1}{36}l_{-1}q_{2}q_{1}q_{-2}\\
& +\frac{1}{36}l_{-1}q_{2}q_{1}q_{-2}-\frac{2}{9}l_{1}q_{1}^{2}q_{-3}+\frac
{1}{36}l_{1}q_{3}q_{-2}^{2}+\frac{1}{9}l_{1}q_{1}q_{-1}^{2}+\frac{1}{18}%
l_{0}q_{3}q_{-1}q_{-2}-\frac{1}{9}l_{-1}q_{1}^{2}q_{-1}\\
& +\frac{1}{36}l_{1}q_{1}q_{0}q_{-2}-\frac{1}{6}l_{0}q_{2}q_{-1}^{2}-\frac
{1}{3}l_{1}q_{3}q_{-1}q_{-3}+\frac{2}{9}l_{-1}q_{3}q_{-1}^{2}-\frac{1}%
{36}l_{-1}q_{2}^{2}q_{-3}-\frac{1}{6}l_{0}q_{1}^{2}q_{-2}\\
& -\frac{1}{6}l_{1}q_{0}^{2}q_{-1}+\frac{1}{6}l_{-1}q_{1}q_{0}^{2}-\frac
{1}{36}l_{-1}q_{2}q_{0}q_{-1}-\frac{1}{36}l_{1}q_{2}q_{-1}q_{-2}+\frac{1}%
{3}l_{-1}q_{3}q_{1}q_{-3}\\
& +\frac{1}{18}l_{0}q_{2}q_{1}q_{-3}-\frac{1}{4}l_{-1}q_{3}q_{0}q_{-2}\\
& \\
\left[  4,0\right]  = & -\frac{1}{9}q_{1}^{3}q_{-3}-\frac{3}{5}q_{3}q_{0}%
^{2}q_{-3}-\frac{1}{36}q_{2}^{2}q_{-1}q_{-3}+\frac{1}{675}q_{2}^{2}q_{-2}%
^{2}+\frac{1}{100}q_{3}q_{2}q_{-2}q_{-3}-\frac{1}{9}q_{3}q_{-1}^{3}\\
& -\frac{1}{15}q_{2}q_{0}^{2}q_{-2}-\frac{5}{108}q_{1}^{2}q_{-1}^{2}-\frac
{1}{540}q_{2}q_{1}q_{-1}q_{-2}+\frac{1}{18}q_{1}^{2}q_{0}q_{-2}+\frac{7}%
{30}q_{3}q_{1}q_{-1}q_{-3}\\
& +\frac{1}{18}q_{2}q_{0}q_{-1}^{2}-\frac{3}{100}q_{3}^{2}q_{-3}^{2}-\frac
{1}{36}q_{3}q_{1}q_{-2}^{2}+\frac{1}{6}q_{2}q_{1}q_{0}q_{-3}+\frac{1}{6}%
q_{3}q_{0}q_{-1}q_{-2}%
\end{array}
\]

A basis of invariants
of $\frak{g}$ can be completed with two additional operators $C_{6}^{\prime}$ and
$C_{8}^{\prime}$ of degrees 6 and 8 respectively. The explicit expression for the
sixth order invariant is
\[
\begin{array}
[c]{ll}%
C_{6}^{\prime}= & -729q_{0}^{6}-54q_{1}^{4}q_{-2}^{2}+54q_{3}q_{-3}\left(
9q_{2}q_{0}^{2}q_{-2}+162q_{1}q_{0}^{2}q_{-1}-32q_{1}^{2}q_{-1}^{2}%
+6q_{2}q_{1}q_{-1}q_{-2}\right)  \\
& +6q_{2}q_{-2}\left(  6q_{3}q_{-1}^{3}-10q_{1}^{2}q_{-1}^{2}+6q_{-3}q_{1}%
^{3}-63q_{1}q_{0}^{2}q_{-1}\right)  -162q_{0}^{2}\left(  q_{-2}^{2}q_{3}%
q_{1}+q_{2}^{2}q_{-3}q_{-1}\right)  \\
& +54\left(  q_{0}^{2}\left(  27q_{3}^{2}q_{-3}^{2}-8q_{-3}q_{1}^{3}%
-8q_{3}q_{-1}^{3}-13q_{1}^{2}q_{-1}^{2}\right)  -q_{3}^{2}\left(  -q_{0}%
q_{-2}^{3}+q_{-1}^{2}q_{-2}^{2}\right)  \right)  \\
& -54\left(  q_{1}^{2}q_{-3}^{2}+q_{-1}^{4}\right)  q_{2}^{2}-3q_{2}^{2}%
q_{-2}^{2}\left(  4q_{1}q_{-1}+9q_{0}^{2}\right)  -324q_{0}^{3}\left(
q_{1}^{2}q_{-2}+q_{2}q_{-1}^{2}\right)  +q_{2}^{3}q_{0}q_{-3}^{2}\\
& -18q_{-2}q_{2}\left(  q_{-2}^{2}q_{3}q_{1}+q_{2}^{2}q_{-3}q_{-1}\right)
-756q_{0}q_{1}q_{-1}\left(  q_{3}q_{-1}q_{-2}+q_{2}q_{1}q_{-3}\right)
-64q_{1}^{3}q_{-1}^{3}\\
& +972\left(  q_{0}^{3}\left(  q_{3}q_{-1}q_{-2}+q_{2}q_{1}q_{-3}\right)
-\left(  q_{3}^{2}q_{-1}q_{-2}q_{-3}+\left(  q_{2}q_{1}q_{-3}^{2}+q_{1}%
^{2}q_{-2}q_{-3}\right)  q_{3}\right)  q_{0}\right)  \\
& +243\left(  6q_{1}q_{-1}-30q_{3}q_{-3}+q_{2}q_{-2}\right)  q_{0}%
^{4}+288q_{-1}q_{1}\left(  q_{-3}q_{1}^{3}+q_{3}q_{-1}^{3}\right)
864q_{-3}^{2}q_{3}q_{1}^{3}\\
& +972q_{2}q_{-1}^{2}q_{-3}q_{3}q_{0}+90q_{-2}q_{2}\left(  q_{1}^{2}%
q_{-2}+q_{2}q_{-1}^{2}\right)  q_{0}+396q_{-1}q_{0}q_{1}\left(  q_{1}%
^{2}q_{-2}+q_{2}q_{-1}^{2}\right)  \\
& +180q_{1}q_{-1}\left(  q_{-2}^{2}q_{3}q_{1}+q_{2}^{2}q_{-3}q_{-1}\right)
+864q_{-3}q_{3}^{2}q_{-1}^{3}+q_{2}^{3}q_{-2}^{3}\\
\end{array}
\]
It turns out, however, that the
invariants $C_{6}^{\prime}$ and $C_{8}^{\prime}$ are not sufficient to provide the labelling operators with the
orthogonality conditions. We have to consider additional operators of degree six which are however not obtained
by contraction\footnote{This is due to the fact that $\frak{so}(5)$ has no primitive Casimir operator
of degree six.}. Denoting the previous $C_{6}^{\prime}$ by $[6,0]$, according to its degree in the $q_{\mu}$ variables,
we introduce the following additional operators of degree four depending on
the parameters $\alpha$ and $\beta$:
\begin{eqnarray}
X^{1}_{1}=
\left[4,0\right]+\left[3,1\right]+\left(4-3\alpha\right)
\left[2,2\right]+\alpha \left[1,3\right],\quad
\alpha\neq 1,\nonumber\\
X_{1}^{2}=\left(\frac{4}{3}-\beta\right)
\left[4,0\right]+\frac{3}{2}\left(1-\beta\right)
\left[3,1\right]+3\beta \left[2,2\right]+\left[1,3\right],\quad
\beta\neq \frac{1}{3}
\end{eqnarray}
and the two operators of degree six:
\begin{equation}
\begin{array}{ll}
X_{2}^{1}=&-\frac{27}{5} \left[6,0\right]-162
\left[5,1\right]+\left[4,2\right]-216\left[3,3\right]-\left[2,0\right]
\left(5310\left[2,2\right]+\frac{2025}{2}\left[4,0\right]\right)\\
 &+\left[0,2\right]\left(2124\left[3,1\right] +528\left[1,3\right]\right)+768\left[2,4\right]\\
 X_{2}^{2}=&-\frac{12}{5} \left[6,0\right]+108
\left[5,1\right]+\left[4,2\right]+324\left[3,3\right]-\left[2,0\right]
\left(180\left[2,2\right]-1035\left[3,1\right]\right) \\
&-\left[0,2\right]\left(17172\left[2,2\right]+1728\left[2,4\right]+3998\left[1,3\right]\right),
\end{array}
\end{equation}

where $[2,4],\; [3,3]\; [4,2]$ and $[5,1]$ are defined in Table 4:

{\small
\begin{sidewaystable}[p]
\caption{Sixth order operators}
\begin{tabular}{|ll|}
\hline
$[2,4]=$ &$(q_{-3}q_{3}-\frac{1}{4}q_{0}^{2})l_{0}^{4}+\frac{1}{2}%
(q_{0}(q_{1}l_{-1}-q_{-1}l_{1})+l_{1}q_{2}q_{-3}-l_{-1}q_{3}q_{-2})l_{0}^{3}+l_{0}^{2}(l_{1}l_{-1}q_{0}^{2}-%
\frac{1}{8}(l_{-1}^{2}q_{2}+l_{1}^{2}q_{-2})q_{0}-\frac{1}{4}l_{1}l_{-1}(-2q_{1}q_{-1}+q_{-2}q_{2}-2q_{-3}q_{3})$ \\
 & $+\frac{1}{4}\left((5q_{3}q_{-1}-q_{1}^{2})l_{-1}^{2}+(+5q_{1}q_{-3}-q_{-1}^{2})l_{1}^{2}\right) )+(\frac{3}{16}
 q_{-1}q_{-3}-\frac{1}{64}q_{-2}^{2})l_{1}^{4}+\frac{1}{32}l_{-1}^{2}l_{1}^{2}(26q_{1}q_{-1}-1q_{-2}q_{2}-50q_{0}^{2}
 +2q_{-3}q_{3})+(\frac{3}{16}q_{3}q_{1}-\frac{1}{64}q_{2}^{2})l_{-1}^{4}$ \\
 & $+\frac{1}{8}l_{1}^{3}((9q_{0}q_{-3}-1q_{-1}q_{-2})l_{0}+\frac{1}{16}(6q_{-1}^{2}+2q_{1}q_{-3}-5q_{0}q_{-2})l_{-1})
 +\frac{1}{16}l_{-1}^{3}((6q_{1}^{2}+2q_{3}q_{-1}-5q_{2}q_{0})l_{1}+\frac{1}{8}(q_{2}q_{1}-9q_{3}q_{0})l_{0})$ \\
 & $+\frac{1}{8}l_{-1}^{2}l_{1}l_{0}(4q_{2}q_{-1}-q_{3}q_{-2}-11q_{1}q_{0})+\frac{1}{8}l_{0}l_{1}^{2}l_{-1}(+q_{2}q_{-3}
 -4q_{1}q_{-2}+11q_{0}q_{-1})$.\\
 \\
$[3,3]=$ &$(q_{2}q_{0}q_{-2}-\frac{4}{3}\left(
q_{1}^{2}q_{-2}-q_{2}q_{-1}^{2}\right)
+8q_{1}q_{0}q_{-1}-9q_{0}^{3})l_{0}^{3}+\frac{1}{12}%
(6q_{0}q_{-1}q_{-2}+12q_{1}q_{-1}q_{-3}-27q_{0}^{2}q_{-3}-4q_{-1}^{3}-q_{1}q_{-2}^{2})l_{1}^{3}+%
\frac{4}{3}\left( q_{1}q_{-1}^{2}-q_{2}q_{-1}q_{-2}\right)
l_{1}l_{0}^{2}$
\\
& $+\frac{1}{12}%
(q_{2}^{2}q_{-1}-12q_{3}q_{1}q_{-1}+4q_{1}^{3}+27q_{3}q_{0}^{2}-6q_{2}q_{1}q_{0})l_{-1}^{3}+(3\left(
q_{2}q_{0}q_{-3}-q_{0}^{2}q_{-1}\right)
+q_{1}q_{0}q_{-2}-4q_{1}^{2}q_{-3})l_{1}l_{0}^{2}+12l_{0}l_{1}^{2}\left(
q_{2}q_{-1}q_{-3}-q_{0}q_{-1}^{2}\right)
$ \\
& $+\frac{1}{3}l_{-1}(12q_{3}q_{-1}^{2}+9\left(
q_{1}q_{0}^{2}-q_{3}q_{0}q_{-2}\right)
+q_{2}q_{1}q_{-2}-3q_{2}q_{0}q_{-1}-4q_{1}^{2}q_{-1})l_{0}^{2}+\frac{1}{12}%
l_{0}(9q_{0}^{2}q_{-2}-36q_{1}q_{0}q_{-3}+4q_{1}q_{-1}q_{-2}-q_{2}q_{-2}^{2})l_{1}^{2}
$ \\
& $+\frac{1}{2}%
l_{0}l_{1}l_{-1}(10q_{1}q_{0}q_{-1}-9q_{0}^{3}-2q_{2}q_{-1}^{2}+2q_{2}q_{1}q_{-3}+2q_{3}q_{-1}q_{-2}-2q_{1}^{2}q_{-2}-18q_{3}q_{0}q_{-3}+q_{2}q_{0}q_{-2})+%
\frac{1}{3}\left(
q_{2}q_{1}q_{-1}+q_{3}q_{1}q_{-2}-q_{1}^{2}q_{0}\right)
l_{-1}^{2}l_{0}$ \\
& $+\frac{1}{12}%
(9q_{2}q_{0}^{2}-36q_{3}q_{0}q_{-1}-q_{2}^{2}q_{-2})l_{-1}^{2}l_{0}-\frac{1}{%
12}%
l_{1}l_{-1}(((8q_{1}^{2}q_{-1}-9q_{1}q_{0}^{2}-36q_{3}q_{1}q_{-3}+18q_{3}q_{0}q_{-2}-2q_{2}q_{1}q_{-2}-12q_{3}q_{-1}^{2}+3q_{2}^{2}q_{-3})l_{-1})
$ \\
& $%
+((36q_{3}q_{-1}q_{-3}+9q_{0}^{2}q_{-1}+2q_{2}q_{-1}q_{-2}-3q_{3}q_{-2}^{2}-8q_{1}q_{-1}^{2}-18q_{2}q_{0}q_{-3}+12q_{1}^{2}q_{-3})l_{1}))
$ \\
&  \\
$[4,2]=$ &$\frac{1}{2}(27l_{1}l_{-1}-135l_{0}^{2})q_{0}^{4}+(27%
\left( l_{0}l_{-1}q_{1}-l_{0}l_{1}q_{-1}\right) -9/2\left(
l_{1}^{2}q_{-2}+l_{-1}^{2}q_{2}\right) )q_{0}^{3}+q_{0}^{2}(l_{0}(\frac{3}{2}%
(l_{1}q_{1}q_{-2}-l_{-1}q_{2}q_{-1})+\frac{27}{2}%
(l_{-1}q_{3}q_{-2}-l_{1}q_{2}q_{-3}))$ \\
& $+(27q_{3}q_{-3}+81q_{1}q_{-1}+9q_{2}q_{-2})l_{0}^{2}+\frac{27}{2}%
l_{-1}(5q_{3}q_{-3}-q_{1}q_{-1})l_{1}+\frac{1}{2}%
(3q_{1}^{2}+9q_{3}q_{-1})l_{-1}^{2}+\frac{1}{2}%
(3q_{-1}^{2}+9q_{1}q_{-3})l_{1}^{2})+16l_{0}^{2}\left(
q_{1}^{3}q_{-3}+q_{3}q_{-1}^{3}\right) $ \\
& $-\frac{1}{6}%
l_{0}^{2}(108(q_{3}q_{0}q_{-1}q_{-2}+q_{2}q_{1}q_{0}q_{-3})-6(q_{3}q_{1}q_{-2}^{2}+q_{2}^{2}q_{-1}q_{-3})+84(q_{1}^{2}q_{0}q_{-2}+q_{2}q_{0}q_{-1}^{2})-20q_{2}q_{1}q_{-1}q_{-2}+q_{2}^{2}q_{-2}^{2}+64q_{1}^{2}q_{-1}^{2})+9l_{1}^{2}q_{3}q_{-3}q_{-1}^{2}
$ \\
& $-\frac{1}{2}l_{1}^{2}q_{2}q_{-2}q_{-1}^{2}-l_{1}^{2}\left( (q_{-1}q_{-3}+%
\frac{3}{4}%
q_{-2}^{2})q_{1}^{2}+(4q_{1}q_{-1}q_{-2}-3q_{2}q_{-1}q_{-3}-9q_{3}q_{-2}q_{-3}+%
\frac{1}{2}%
q_{2}q_{-2}^{2})q_{0}-q_{1}q_{-1}^{3}+9q_{3}q_{1}q_{-3}^{2}+q_{2}q_{1}q_{-2}q_{-3}-%
\frac{3}{4}q_{2}^{2}q_{-3}^{2}\right) $ \\
& $+l_{-1}^{2}((-9q_{3}q_{2}q_{-3}-3q_{3}q_{1}q_{-2}+\frac{1}{2}%
q_{2}^{2}q_{-2}+4q_{2}q_{1}q_{-1})q_{0}+(9q_{3}q_{-3}-\frac{1}{2}%
q_{2}q_{-2})q_{1}^{2}-(q_{3}q_{1}+\frac{3}{4}%
q_{2}^{2})q_{-1}^{2}+9q_{3}^{2}q_{-1}q_{-3}-q_{1}^{3}q_{-1}+q_{3}q_{2}q_{-1}q_{-2}-%
\frac{3}{4}q_{3}^{2}q_{-2}^{2})$ \\
& $%
+(-6l_{0}l_{1}q_{2}+4l_{1}l_{-1}q_{3})q_{-1}^{3}+(6l_{0}l_{-1}q_{-2}+4l_{1}l_{-1}q_{-3})q_{1}^{3}+q_{1}^{2}(l_{0}((-28q_{-1}l_{-1}+12l_{1}q_{-3})q_{0}-(6l_{-1}q_{2}q_{-3}+8l_{1}q_{-1}q_{-2}))+%
\frac{1}{3}l_{-1}l_{1}(3q_{0}q_{-2}+4q_{-1}^{2}))$ \\
& $%
+q_{-1}^{2}((-12l_{0}l_{-1}q_{3}+l_{1}l_{-1}q_{2}+28l_{0}l_{1}q_{1})q_{0}+(8l_{-1}q_{2}q_{1}+6q_{-2}q_{3}l_{1})l_{0})+l_{1}l_{-1}(%
\frac{5}{2}q_{3}q_{1}q_{-2}^{2}-9q_{3}q_{0}q_{-1}q_{-2}+\frac{1}{3}%
q_{2}q_{1}q_{-1}q_{-2}+\frac{5}{2}q_{2}^{2}q_{-1}q_{-3}-\frac{1}{6}%
q_{2}^{2}q_{-2}^{2}$ \\
& $%
-36q_{3}q_{1}q_{-1}q_{-3}-9q_{2}q_{1}q_{0}q_{-3})+l_{0}l_{1}((6l_{-1}q_{2}^{2}q_{-3}+5l_{1}q_{2}q_{-1}q_{-2}-6l_{1}q_{3}q_{-2}^{2}-5l_{-1}q_{2}q_{1}q_{-2})q_{0}+(6q_{3}q_{1}q_{-2}q_{-3}+%
\frac{1}{2}\left( q_{2}q_{1}q_{-2}^{2}-q_{2}^{2}q_{-2}q_{-3}\right) )$ \\
&
$+2l_{0}l_{1}q_{2}q_{1}q_{-1}q_{-3}+l_{0}l_{-1}(\frac{1}{2}\left(
q_{3}q_{2}q_{-2}^{2}-q_{2}^{2}q_{-1}q_{-2}\right)
-6q_{3}q_{2}q_{-1}q_{-3}-2q_{3}q_{1}q_{-1}q_{-2}).$ \\
 & \\
$[5,1]=$ & $-8l_{0}q_{0}^{5}-\frac{4}{3}%
(l_{1}q_{-1}-l_{-1}q_{1})q_{0}^{4}+\frac{2}{9}%
q_{0}^{3}((2q_{2}q_{-2}-9q_{3}q_{-3}+58q_{1}q_{-1})l_{0}+(6q_{2}q_{-3}+q_{1}q_{-2})l_{1}-(q_{2}q_{-1}+6q_{3}q_{-2})l_{-1})+q_{2}^{2}q_{-3}l_{1}(%
\frac{1}{12}q_{3}q_{-3}-\frac{1}{81}q_{2}q_{-2})$ \\
& $+\frac{43}{81}(l_{-1}q_{1}^{3}q_{-1}^{2}-l_{1}q_{1}^{2}q_{-1}^{3})-\frac{1%
}{108}(l_{0}q_{3}^{2}q_{-2}^{3}+l_{0}q_{2}^{3}q_{-3}^{2})+q_{0}^{2}(\frac{5}{%
27}\left( l_{-1}q_{2}q_{1}q_{-2}-l_{1}q_{2}q_{-1}q_{-2}\right) +\frac{1}{3}%
\left(
l_{1}q_{3}q_{-2}^{2}-l_{-1}q_{2}^{2}q_{-3}+l_{0}q_{2}q_{1}q_{-3}+l_{0}q_{3}q_{-1}q_{-2}\right)
)
$ \\
& $+\left( \frac{11}{9}\left(
l_{-1}q_{3}q_{-1}^{2}-l_{1}q_{1}^{2}q_{-3}\right)
+\frac{5}{3}\left(
l_{-1}q_{3}q_{1}q_{-3}-l_{1}q_{3}q_{-1}q_{-3}\right)
+\frac{16}{9}\left(
l_{1}q_{1}q_{-1}^{2}-l_{-1}q_{1}^{2}q_{-1}\right)
-\frac{38}{27}\left(
l_{0}q_{2}q_{-1}^{2}+l_{0}q_{1}^{2}q_{-2}\right) )\right) q_{0}^{2}+\frac{1}{%
36}(l_{1}q_{1}^{3}q_{-2}^{2}-l_{-1}q_{2}^{2}q_{-1}^{3})$ \\
& $-\frac{1}{18}%
(l_{1}q_{3}q_{2}q_{-1}q_{-2}q_{-3}-l_{-1}q_{3}q_{2}q_{1}q_{-2}q_{-3})+\frac{4%
}{27}(l_{1}q_{3}q_{-1}^{4}-l_{-1}q_{1}^{4}q_{-3})+l_{0}q_{0}(\frac{4}{27}%
\left( q_{1}^{3}q_{-3}+q_{3}q_{-1}^{3}\right) +\frac{1}{18}\left(
q_{3}q_{1}q_{-2}^{2}+q_{2}^{2}q_{-1}q_{-3}\right)
+q_{3}^{2}q_{-3}^{2}(l_{1}q_{-1}-l_{-1}q_{1})$ \\
& $+\frac{10}{81}q_{1}q_{-1}(l_{-1}q_{2}^{2}q_{-3}-l_{1}q_{3}q_{-2}^{2})+%
\frac{1}{18}l_{0}q_{0}(20q_{1}q_{-1}-11q_{2}q_{-2}+18q_{3}q_{-3})q_{3}q_{-3}+%
\frac{4}{81}q_{2}^{2}q_{-2}^{2}l_{0}q_{0}-\left( \frac{403}{81}%
q_{1}^{2}q_{-1}^{2}+\frac{5}{54}q_{2}q_{1}q_{-1}q_{-2}\right) l_{0}q_{0}+%
\frac{68}{81}q_{1}^{3}q_{-3}q_{-1}l_{1}$ \\
& $+(\frac{8}{9}q_{1}q_{-1}-\frac{1}{12}q_{-2}q_{2}+\frac{8}{27}%
q_{3}q_{-3})(q_{-1}^{2}q_{2}+q_{1}^{2}q_{-2})l_{0}+(\frac{7}{162}q_{-2}q_{2}-%
\frac{32}{81}q_{1}q_{-1})(q_{3}q_{-1}q_{-2}+q_{1}q_{-3}q_{2})l_{0}+\frac{19}{%
162}q_{2}q_{-2}(l_{-1}q_{3}q_{-1}^{2}-l_{1}q_{-3}q_{1}^{2})$ \\
& $+q_{0}(\frac{41}{162}q_{1}q_{-1}+\frac{7}{18}q_{3}q_{-3}+\frac{2}{81}%
q_{-2}q_{2})(q_{2}l_{-1}q_{-1}-l_{1}q_{-2}q_{1})+q_{0}(\frac{4}{27}%
q_{-2}q_{2}-\frac{47}{54}q_{1}q_{-1}-\frac{5}{6}%
q_{3}q_{-3})(q_{-3}q_{2}l_{1}-l_{-1}q_{3}q_{-2})-\frac{2}{27}%
l_{0}l_{1}q_{-1}^{2}(q_{-1}q_{2}+4q_{-2}q_{3})$ \\
& $+\frac{2}{27}q_{0}q_{1}^{2}l_{-1}(q_{1}q_{-2}+4q_{-3}q_{2})-(\frac{38}{27}%
q_{-3}q_{3}q_{-1}q_{1}+\frac{17}{81}q_{1}q_{-1}q_{-2}q_{2}-\frac{1}{81}%
q_{2}^{2}q_{-2}^{2})(q_{1}l_{-1}-l_{1}q_{-1})+l_{1}q_{1}q_{-3}^{2}(\frac{8}{9%
}q_{3}q_{1}-\frac{1}{108}q_{2}^{2})+l_{1}q_{-3}q_{-1}(\frac{68}{81}q_{1}^{3}-%
\frac{5}{108}q_{2}^{2}q_{-1})$ \\
& $-\frac{5}{108}l_{1}q_{-3}q_{-1}^{2}q_{2}^{2}+(\frac{1}{81}q_{2}q_{-2}-%
\frac{1}{12}q_{3}q_{-3})q_{3}q_{-2}^{2}l_{1}+l_{-1}q_{3}^{2}q_{-1}(\frac{1}{%
108}q_{-2}^{2}-\frac{8}{9}q_{-1}q_{-3})+l_{-1}q_{1}q_{3}(\frac{5}{108}%
q_{1}q_{-2}^{2}-\frac{68}{81}q_{-1}^{3})$ \\
& \\\hline
\end{tabular}
\end{sidewaystable}
}

Observe that for the excluded values of the parameters, we recover the Casimir operator of
degree four of $\frak{so}(5)$. We claim that the operators
$\left[0,2\right],\left[2,0\right],C_{4},X_{1}^{1},X_{1}^{2},X_{2}^{1},X_{2}^{2}$
are functionally independent, where $[0,2]=C_{21}$ and $[2,0]=C_{2}-C_{21}$. To prove this,
we simply consider the following Jacobian
\begin{equation}
\frac{\partial\left(\left[0,2\right],\left[2,0\right],C_{4},X_{1}^{1},X_{1}^{2},X_{2}^{1},X_{2}^{2}\right)}
{\partial\left(l_{0},l_{-1},q_{0},q_{-1},q_{-2},q_{-3},q_{3}\right)}\neq
0
\end{equation}
Therefore the operators are independent, and from them a set of commuting operators can be extracted.

\begin{proposition}
The sets
$\mathcal{F}_{1,\alpha}=\left\{X_{1}^{1},X_{2}^{1}\right\}$ and
$\mathcal{F}_{2,\beta}=\left\{X_{1}^{2},X_{2}^{2}\right\}$ are
inequivalent set of commuting missing label operators.
\end{proposition}

The non-equivalence of the sets of labelling operators refer to their independence and to the
fact they they are not mutually orthogonal. This shows that the class of labelling operators is
divided into two types with respect to the orthogonality requirement. It cannot however excluded
that some mixed functions can have the same property.

\smallskip

We finally remark that the solution found in \cite{Me} is equivalent to
the symmetrized solution $\left\{X_{1}^{1},X_{1}^{2}\right\}$ for $\alpha=0$.
In fact, we obtain $X^{\prime}=X_{1}^{1}-C_{4}=3\left[2,2\right]-\left[1,3\right]$,
therefore the only solution with two components. In this sense, the pair proposed
in \cite{Me} is actually the simplest possible choice for solving the missing label
problem.

\section{Cartan subalgebras and the MLP}

As already observed, the eigenvalues of the Casimir operators of a (semisimple) Lie
algebra serve to characterize the irreducible representations. The next step is to
find out how many internal labels are necessary to distinguish the different eigenvectors
of a given weight. Obviously the eigenvalues of the Cartan subalgebra generators serve to this
purpose, but with the exception of the Lie algebras $[\frak{su}(2)]^{n}$, they are not sufficient.
It is well known from the classical theory that for (semisimple) algebras, exactly
$\frac{1}{2}\left(\dim\frak{s}-{\rm rank}\; \frak{s}\right)$ labels are needed.\footnote{This is exactly the
number of positive roots. Using the Maurer-Cartan equations of the algebra, it can be shown \cite{C43} that
this number appears naturally}. This means that in addition to the generators of the Cartan subalgebra $\frak{h}$, we
need to find $f=\frac{1}{2}\left(\dim\frak{s}-3{\rm rank} \frak{s}\right)$ additional operators
to separate multiplicities. These must of course commute among themselves and with the elements of the Cartan
subalgebra. The number $f$ is usually called the Racah number, and the corresponding operators the
Racah operators. As follows from the expression of $f$, this number increases quickly for high dimensions, and so, for
example, for the exceptional algebra $E_{8}$ we need as many as $112$ additional operators. The problem to completely
characterize the Racah operators is still open, and only for some special cases it has been studied. The problem is
however quite similar to that of the MLP, and this suggest to emply the same technique to look for these labelling
operators. The main difference is that the reduction chain $\frak{s}\supset \frak{h}$ involves an Abelian subalgebra,
and therefore the contraction will generally be not an inhomogeneous Lie algebra. In any case, the decomposition of
the Casimir operators of $\frak{s}$ provides solutions to the problem, since they are particular solutions to the
system of differential equations determined by the Cartan subalgebra. A suggested procedure to obtain Racah operators
is therefore the following:

\begin{enumerate}

\item Decompose the Casimir operators of $\frak{s}$ according to the contraction associated to the reduction
$\frak{s}\supset \frak{h}$.

\item Extract a maximal number of independent operators that are moreover independent on the generators of $\frak{h}$
and the Casimir operators of $\frak{s}$

\item Compute the brackets of these operators in order to obtain a set of commuting operators.
\end{enumerate}

Observe in particular that since the Racah operators are solutions to the corresponding differential operators, commutation
with the Cartan subalgebra generators follows at once. Once the brackets of the Racah operators between themselves and
the Casimir operators of the algebra must be computed. It is clear from this procedure that an exact knowledge of the Casimir operators of simple Lie algebras is necessary.
Formulae for these are well known \cite{Pe}, although in some cases the corresponding expressions are not
very manageable. Specially for the $E_{i}$ series, where the higher order Casimir operators are quite difficult
to compute, this approach would require alternative derivations of the invariants, as done for the classical series.
The question {\it how many} of the Racah operators can be obtained using this method is still an open problem,
actually in progress.

\medskip

In this paragraph, we show that this ansatz works, even if the nature of the corresponding missing label problem
is quite different to the usual one, due to the non-inhomogeneous nature of the contracted algebra. The examples
exhibited point out that the interpretation of labelling operators as ``broken Casimir operators" is valid, and
gives to the Racah operators a certain physical meaning absent in other approaches to their computation.

\subsection{Racah operators of  $\frak{su}(3)$}

The lowest dimensional simple algebra where such operators are required is the type $A_{2}$. In this
case, $f=1$, thus we need to determine one additional operator. As follows from (\ref{ML}), there are
two possibilities. For simplicity, we consider a basis $\left\{  L_{i},T_{j}\right\} $ $(i=0,1,-1,\; j=-2,..,2)$
similar to that
considered for the Elliott chain, but changing the indices for the second rank tensor. The commutator table
in matrix form is given by:
\begin{equation*}
A(\frak{su}(3))=\left(
\begin{array}{cccccccc}
0 & {l_{1}} &  - {l_{-1}} & 2\,{t_{2}} & {t_{1}} & 0 &  - {t_{-1}
} &  - 2\,{t_{-2}} \\
 - {l_{1}} & 0 & {l_{0}} & 0 &  - 2\,{t_{2}} &  - 3\,{t_{1}} & {t
_{0}} & {t_{-1}} \\
{l_{-1}} &  - {l_{0}} & 0 &  - {t_{1}} &  - {t_{0}} & 3\,{t_{-1}}
 & 2\,{t_{-2}} & 0 \\
 - 2\,{t_{2}} & 0 & {t_{1}} & 0 & 0 & 0 &  - {l_{1}} & {l_{0}} \\
 - {t_{1}} & 2\,{t_{2}} & {t_{0}} & 0 & 0 &  - 3\,{l_{1}} & {l_{0
}} &  - {l_{-1}} \\
0 & 3\,{t_{1}} &  - 3\,{t_{-1}} & 0 & 3\,{l_{1}} & 0 &  - 3\,{l_{
-1}} & 0 \\
{t_{-1}} &  - {t_{0}} &  - 2\,{t_{-2}} & {l_{1}} &  - {l_{0}} & 3
\,{l_{-1}} & 0 & 0 \\
2\,{t_{-2}} &  - {t_{-1}} & 0 &  - {l_{0}} & {l_{-1}} & 0 & 0 & 0
\end{array}
\right)
\end{equation*}
In this case, the Cartan subalgebra is easily seen to be generated by $T_{0}$ and $L_{0}$.
To compute the Racah operators, we have to solve the system:
\begin{equation}
\begin{array}{l}
\widehat{T}_{0}F:=l_{1}\frac{\partial F}{\partial l_{1}}-l_{-1}\frac{\partial
F}{\partial l_{-1}}+2t_{2} \frac{\partial F}{\partial
t_{2}}+t_{1}\frac{\partial F}{\partial t_{1}}-t_{-1}\frac{\partial
F}{\partial t_{-1}}-2t_{-2}\frac{\partial F}{\partial
t_{-2}}=0,\\
\widehat{T}_{0}F:=3t_{1}\frac{\partial F}{\partial
l_{1}}-3t_{-1}\frac{\partial F}{\partial
l_{-1}}+3l_{1}\frac{\partial F}{\partial
t_{1}}-3l_{-1}\frac{\partial F}{\partial t_{-1}}=0.
\end{array}
\end{equation}

Instead of integrating the system, we decompose the Casimir operators. Over this basis, the
unsymmetrized invariants are $C^{2}=2 l_{1}l_{-1}+l_{0}^2+4 t_{2}t_{-2}+2 t_{1}t_{-1}+\frac{1}{3}t_{0}^2$
and
\begin{equation*}
\begin{array}{ll}
C^{3}=&9\left(l_{1}l_{-1}t_{0} - l_{0}^2t_{0}+ t_{1}t_{0}t_{-1}\right)-36 t_{ 2}t_{0}t_{-2}+t_{0}^3\\
& +27\left(t_{2}t_{-1}^2+t_{1}^2t_{-2}- l_{0}l_{1}t_{-1}- l_{0}l_{-1}t_{1}- l_{1}^2t_{-2}- l_{-1}^2t_{2}\right).
\end{array}
\end{equation*}

Only the decomposition of the cubic operator can lead to an independent labelling operator. In this case, the
contraction is determined by the transformations
\[
L^{\prime}_{0}=L_{0}; L^{\prime}_{i}=\frac{1}{t}L_{i},\quad i=\pm 1;\;
T^{\prime}_{0}=T_{0}; T^{\prime}_{i}=\frac{1}{t}T_{i},\quad i=\pm 1,2.
\]
The re-scaled cubic operator is therefore
\begin{equation*}
\begin{array}{ll}
C^{3}=&\left(t_{0}^{3}-9l_{0}^{2}t_{0}\right)+t^{2}\left(9\left(l_{1}t_{0}l_{-1}+t_{1}l_{0}t_{-1}\right)
-27\left(t_{-1}l_{0}l_{1}+l_{-1}l_{0}t_{-1}\right)-36t_{0}t_{2}t_{-2}\right)\\
&-t^{3}\left(l_{1}^{2}t_{-2}+l_{-1}^{2}t_{2}-t_{2}t_{-1}^{2}-t_{1}^{2}t_{-2}\right).
\end{array}
\end{equation*}
The first term is obviously non useful, for being a function of the Cartan generators. Choosing
$\Phi=9\left(l_{1}t_{0}l_{-1}+t_{1}l_{0}t_{-1}\right)
-27\left(t_{-1}l_{0}l_{1}+l_{-1}l_{0}t_{-1}\right)-36t_{0}t_{2}t_{-2}$ provides an independent labelling
operator, as follows from the Jacobian
\begin{equation}
\frac{\partial\left\{  T_{0},L_{0},C^{2},C^{3},\Phi\right\}  }{\partial
\left\{  l_{0},t_{0},t_{2},t_{1},t_{-1}\right\}  }\neq0.
\end{equation}
The orthogonality is straightforward. Therefore the symmetrization of
 $\Phi$ constitutes an admissible Racah operator
for the Lie algebra $\frak{su}(3)$.

\subsection{Racah operators of  $\frak{sp}(4)$}

Symplectic groups naturally appear in physical applications from
the boson formalism: the generators of the corresponding Lie algebra
correspond to operators changing the number of particles. In the study
of the $j-j$-coupling shell model, the algebra $\frak{sp}(4)$ played
a notorious role, since they commute with the quasi-spin operators and
moreover cannot change the seniority number $\nu$. More recently, they have
been shown to be important in the nuclear collective model, among other
applications to nuclear physics \cite{Hel,Dra}. Since the symplectic algebra
$\frak{sp}(4)$ has dimension ten and rank two, the Racah number is $f=2$, and
therefore we have to find two commuting labelling operators. Using the same basis
and brackets of (\ref{Kl3}), i.e., the Racah realization of the algebra, where
the indices run from $-2$ to $2$, the Casimir operators can be computed using the matrix
formula (\ref{M0}) adapted to it. The unsymmetrized invariants are
\[%
\begin{array}
[c]{cl}%
C_{2}= & -x_{1,1}^{2}-x_{2,-2}x_{-2,2}-x_{2,2}^{2}-2x_{-1,2}x_{1,-2}%
-x_{1,-1}x_{-1,1}-2x_{2,1}x_{1,2},\\
C_{4}= & 2x_{1,1}x_{-1,2}x_{1,-2}x_{2,2}+x_{1,-1}x_{-1,1}x_{2,2}^{2}%
+x_{2,-2}x_{-2,2}x_{1,1}\symbol{94}2-2x_{1,1}x_{2,2}x_{2,1}x_{1,2}\\
& -2x_{1,-1}x_{2,1}x_{-1,2}x_{2,2}-2x_{1,1}x_{1,-2}x_{-2,2}x_{2,1}%
-2x_{1,1}x_{2,-2}x_{-1,2}x_{1,2}+x_{-1,2}^{2}x_{1,-2}^{2}\\
& -2x_{1,2}x_{1,-2}x_{-1,1}x_{2,2}+x_{2,1}^{2}x_{1,2}^{2}-x_{1,-2}^{2}%
x_{-1,1}x_{-2,2}-x_{1,-1}x_{2,-2}x_{-1,2}^{2}+x_{1,1}^{2}x_{2,2}^{2}\\
& +x_{1,-1}x_{-2,2}x_{2,1}^{2}+2x_{2,1}x_{-1,2}x_{1,-2}x_{1,2}+x_{1,-1}%
x_{2,-2}x_{-1,1}x_{-2,2}+x_{2,-2}x_{-1,1}x_{1,2}^{2}.
\end{array}
\]
Taking the contraction defined by the transformations $X_{i,j}^{\prime}=\frac{1}{t}X_{i,j}$ for
all indices ${i,j\;|i\neq j}$, the quartic Casimir operator is decomposed as:
\begin{equation}
C_{4}=x_{1,1}^{2}x_{2,2}^{2}+t^{2}C_{[2,2]}+t^{3}C_{[1,3]}+t^{4}C_{[0,4]},
\end{equation}
where
\[%
\begin{array}
[c]{cl}%
C_{\left[  2,2\right]  }= & \ 2x_{1,1}x_{-1,2}x_{1,-2}x_{2,2}+x_{1,-1}%
x_{-1,1}x_{2,2}\symbol{94}2+x_{2,-2}x_{-2,2}x_{1,1}^{2}-2x_{1,1}x_{2,2}%
x_{2,1}x_{1,2},\\
C_{\left[  1,3\right]  }= & -2x_{1,-1}x_{2,1}x_{-1,2}x_{2,2}-2x_{1,1}%
x_{1,-2}x_{-2,2}x_{2,1}-2x_{1,1}x_{2,-2}x_{-1,2}x_{1,2}\\
& -2x_{1,2}x_{1,-2}x_{-1,1}x_{2,2}\\
C_{\left[  0,4\right]  }= & x_{2,1}^{2}x_{1,2}^{2}-x_{1,-2}^{2}x_{-1,1}%
x_{-2,2}-x_{1,-1}x_{2,-2}x_{-1,2}^{2}+x_{-1,2}^{2}x_{1,-2}^{2}+x_{1,-1}%
x_{-2,2}x_{2,1}^{2}\\
& +2x_{2,1}x_{-1,2}x_{1,-2}x_{1,2}+x_{1,1}^{2}x_{2,2}^{2}+x_{1,-1}%
x_{2,-2}x_{-1,1}x_{-2,2}+x_{2,-2}x_{-1,1}x_{1,2}^{2}%
\end{array}
\]
Taking for example the operators $C_{[2,2]}$ and $C_{[1,3]}$, we verify their
independence on the Casimir operators of $\frak{sp}(4)$ and the Cartan generators by
computing the Jacobian:
\[
\frac{\partial\left\{  X_{1,1},X_{2,2},C_{2},C_{4},C_{[2,2]},C_{[1,3]}%
\right\}  }{\partial\left\{  x_{1,1},x_{2,2},x_{1,2},x_{2,1},x_{1,-1}%
,x_{-1,2}\right\}  }\neq0.
\]
Taking the symmetrization of these operators (denoted by the same symbol), we finally
compute the brackets:
\begin{equation}
\begin{array}
[c]{lll}%
\left[  C_{[2,2]},C_{[1,3]}\right]  =0, & \left[  C_{[2,2]},C_{2}\right]
=0, & \left[  C_{[2,2]},C_{4}\right]  =0,\\
\left[  C_{[1,3]},C_{2}\right]  =0, & \left[  C_{4},C_{[1,3]}\right]  =0. &
\end{array}
\end{equation}
This shows that the commuting Racah operators also arise as ``broken Casimir operators"

\medskip

The symplectic algebra $\frak{sp}\left(  6\right)  $ shows more
clearly to which extent the described procedure is valid. In this
case, the Racah number equals $f=6$, thus we have to determine six
commuting labelling operators. Again, taking the realization used
before, the invariants follow from formula (\ref{LOL}). Before
symmetrization, the Casimir operators $C_{2},C_{4}$ and $C_{6}$
have $12,123$ and $388$ terms, respectively. Performing the
contraction determined by $X_{i,j}^{\prime}=\frac{1}{t}X_{i,j}$
for $i\neq j$, the quartic and hexic Casimir operators decompose
as
\begin{equation}
\begin{array}{ll}
C_{4}  & =C_{\left[  4,0\right]  }+{t^{2}}C_{\left[  2,2\right]
}+{t^{3}}C_{\left[  1,3\right]  }+{t^{4}}C_{\left[
0,4\right]  },\\
C_{6}  & =C_{\left[  6,0\right]  }+{t^{2}}C_{\left[  4,2\right]
}+{t^{3}}C_{\left[  3,3\right]  }+{t^{4}}C_{\left[
2,4\right]  }+{t^{5}}C_{\left[  1,5\right]  }+{t^{6}%
}C_{\left[  0,6\right]  }.
\end{array}
\end{equation}
Here $C_{\left[  4,0\right]  }$ and $C_{\left[  6,0\right]  }$ are functions
of the Cartan generators, while $C_{\left[  2,2\right]  },C_{\left[
1,3\right]  },C_{\left[  0,4\right]  },C_{\left[  4,2\right]  },C_{\left[
3,3\right]  },C_{\left[  2,4\right]  },C_{\left[  1,5\right]  }$ and
$C_{\left[  0,6\right]  }$ are polynomials with $18,36,66,9,20,72,132$ and
$154$ terms, respectively.\footnote{For this reason we omit the explicit
expressions of the operators.} Taking the six independent operators
$C_{\left[  2,2\right]  },C_{\left[  1,3\right]  },C_{\left[  4,2\right]
},C_{\left[  3,3\right]  },C_{\left[  2,4\right]  }$ and $C_{\left[
1,5\right]  }$, their independence with the $C_{i}$ and the Cartan subalgebra
is proved by means of the Jacobian
\[
\frac{\partial\left\{  X_{1,1},X_{2,2},X_{3,3},C_{2},C_{4},C_{6},C_{\left[
2,2\right]  },C_{\left[  1,3\right]  },C_{\left[  4,2\right]  },C_{\left[
3,3\right]  },C_{\left[  2,4\right]  },C_{[1,5]}\right\}  }{\partial\left\{
x_{1,1},x_{2,2},x_{3,3},x_{1,2},x_{1,3},x_{2,1},x_{2,3},x_{-1,2}%
,x_{-1,3},x_{1,-1},x_{-1,2},x_{2,-3}\right\}  }\neq0.
\]
A straightforward but tedious computation shows moreover the orthogonality
conditions required:%
\begin{equation}
\begin{array}
[c]{ccc}%
\left[  C_{[2,2]},C_{[1,3]}\right]  =0, & \left[  C_{[2,2]},C_{[4,2]}\right]
=0, & \left[  C_{[2,2]},C_{[3,3]}\right]  =0,\\
\left[  C_{[2,2]},C_{[2,4]}\right]  =0, & \left[  C_{[2,2]},C_{[1,5]}\right]
=0, & \left[  C_{[1,3]},C_{[4,2]}\right]  =0,\\
\left[  C_{[1,3]},C_{[3,3]}\right]  =0, & \left[  C_{[1,3]},C_{[2,4]}\right]
=0, & \left[  C_{[1,3]},C_{[1,5]}\right]  =0,\\
\left[  C_{[4,2]},C_{[3,3]}\right]  =0, & \left[  C_{[4,2]},C_{[2,4]}\right]
=0, & \left[  C_{[4,2]},C_{[1,5]}\right]  =0,\\
\left[  C_{[3,3]},C_{[2,4]}\right]  =0, & \left[  C_{[3,3]},C_{[1,5]}\right]
=0, & \left[  C_{[2,4]},C_{[1,5]}\right]  =0,
\end{array}
\end{equation}
Orthogonality with the Casimir operators follows similarly. This suggest that a
complete characterization of the Racah operators by means of the decomposition
of the Casimir operators is possible. This problem is in progress.

\subsection{Applications to the Lie algebra $\frak{so}(2,4)$}

The conformal
group $SO(2,4)$ is one of the physically most relevant groups, and appears in
many different context. One one hand, it is the symmetry group of the Maxwell
equations of electromagnetism, as well as the dynamical
non-invariance group of hydrogen-like atoms. In General Relativity, this group
and their De Sitter subgroups also constitute a powerful tool. This group was
also found to be at the basis of the group-theoretical construction of the Periodic
Table of chemical Elements, being first considered in \cite{Bar}. various authors
followed the analysis of the chemical elements using the conformal symmetry \cite{Ru}.
It is nowadays at the centre of the program KGR \cite{Ki}, whose finality is to obtain
quantitative predictions on the elements \cite{Ki,Ki2}. The Casimir and Racah operators
are the main tool to construct the needed quantum numbers to characterize physical and chemical
properties. Since the Lie algebra is of dimension 15 and rank 3, we need three
additional operators to obtain a complete set of commuting operators
that solves the labelling problem. Once again we try to solve it via the MLP
associated to the Cartan subalgebra. We use the important fact that $\frak{so}(2,4)$ is
isomorphic to the unitary Lie algebra $\frak{su}(2,2)$. Taking the basis formed by the operators $\left\{
E_{\mu\nu},F_{\mu\nu}\right\}  _{1\leq\mu,\nu\leq p+q=n}$ with the constraints%
\begin{eqnarray}
E_{\mu\nu}+E_{\nu\mu}=0,\;F_{\mu\nu}-F_{\nu\mu}=0,\nonumber\\
g_{\mu\mu}=(\left(1,1,-1,-1\right),\nonumber
\end{eqnarray}
the brackets are then given by
\begin{eqnarray}
\left[  E_{\mu\nu},E_{\lambda\sigma}\right]   &
=g_{\mu\lambda}E_{\nu\sigma
}+g_{\mu\sigma}E_{\lambda\nu}-g_{\nu\lambda}E_{\mu\sigma}-g_{\nu\sigma
}E_{\lambda\mu}\label{b1}\nonumber\\
\left[  E_{\mu\nu},F_{\lambda\sigma}\right]   &
=g_{\mu\lambda}F_{\nu\sigma
}+g_{\mu\sigma}F_{\lambda\nu}-g_{\nu\lambda}F_{\mu\sigma}-g_{\nu\sigma
}F_{\lambda\mu}\nonumber\\
\left[  F_{\mu\nu},F_{\lambda\sigma}\right]   &
=g_{\mu\lambda}E_{\nu\sigma
}+g_{\nu\lambda}E_{\mu\sigma}-g_{\nu\sigma}E_{\lambda\mu}-g_{\mu\sigma
}E_{\lambda\nu}\label{b21}%
\end{eqnarray}
To recover the conformal algebra, we take the Cartan subalgebra
spanned by the vectors
$H_{\mu}=g_{\mu+1,\mu+1}F_{\mu\mu}-g_{\mu\mu}F_{\mu+1,\mu+1}$ for
$\mu=1..3$. The centre of $\frak{u}(p,q)$ is obviously generated
by $g^{\mu\mu}F_{\mu\mu}$.\footnote{Another useful basis is that spanned by the
so-called Yao generators.} The advantage of this basis is that the Casimir operators
can be determined by means of a determinantal formula. Indeed, a
 maximal set of independent Casimir invariants of
$\frak{su}\left(  2,2\right)  $ is given by the coefficients
$C_{k}$ of the characteristic polynomial $\left| I A-\lambda {\rm
Id}_{N}\right| =\lambda^{4}+\sum_{k=2}^{4}D_{k}\lambda^{4-k}$,
where $A$ is the matrix defined by:
\begin{equation}
\left(
\begin{array}
[c]{ccccc}%
-I(\frac{3}{4}h_{1}-\frac{1}{2}h_{2}+\frac{1}{4}h_{3}) &-e_{12}-I
f_{12} & e_{13}+I f_{13} & e_{14}+I f_{14} \\
e_{12}-I f_{12} & I (\frac{1}{4} h_{1}+\frac{1}{2}
h_{2}-\frac{1}{4} h_{3}) & e_{23}+I f_{23} & e_{24}+I
f_{24}\\
e_{13}-I f_{13} & e_{23}-I f_{23} & I (\frac{1}{4}
h_{1}-\frac{1}{2} h_{2}-\frac{1}{4} h_{3}) & e_{34}+I f_{34}\\
e_{14} -I f_{14} & e_{24}-I f_{24} &-e_{34}+I f_{34} & I
(\frac{1}{4}h_{1}-\frac{1}{2}h_{2}+\frac{3}{4}h_{3})
\end{array}
\right). \label{Tv}
\end{equation}

The Casimir operators follow from the corresponding symmetrization of the
functions $C_{k}$. The corresponding
contraction is defined by the non-singular transformations
\[
E^{\prime}_{ij}=\frac{1}{t}E_{ij},\;F^{\prime}_{ij}=\frac{1}{t}F_{ij},\quad i=1..4.
\]
According to this scheme, the Casimir operators decompose as follows:
\begin{equation}%
\begin{array}
[c]{l}%
C_{2}=C_{\left[  2,0\right]  }+{t^{2}}C_{\left[  (,2\right]  },\\
C_{3}=C_{[3,0]}+{t^{2}}C_{[1,2]}+{t^{3}}C_{[0,3]},\\
C_{4}=C_{[4,0]}+{t}C_{[1,3]}+{t^{2}}C_{[2,2]}+{t^{4}%
}C_{[0,4]},
\end{array}
\end{equation}
where the $C_{[0,i]}$ are functions of the Cartan generators. The functions
$C_{\left[  i,j\right]  }$ all constitute solutions to the MLP. To complete
the set of orthogonal operators $\left\{  H_{1},H_{2},H_{3},C_{2},C_{3}%
,C_{4}\right\}  $ with three mutually commuting labelling operators, we chose
those triples that are functionally independent from the Casimir operators of
$\frak{su}(2,2)$ and the $h_{i}$. We can take for example $C_{[2,1]}%
,C_{[3,1]}$ and $C_{[2,2]}$. Since
\begin{equation}
\frac{\partial{\left(  H_{1},H_{2},H_{3},C_{2},C_{3},C_{4},C_{\left[
2,1\right]  },C_{\left[  1,3\right]  },C_{\left[  2,2\right]  }\right)  }%
}{\partial{\left(  h_{1},h_{2},h_{3},e_{12},e_{13},e_{14},f_{23},f_{24}%
,f_{34}\right)  }}\neq0,
\end{equation}
these operators are independent. A laborious computation shows that the
symmetrized forms of these operators are orthogonal:%
\begin{equation}
\begin{array}
[c]{lll}%
\left[  C_{i},C_{\left[  1,2\right]  }\right]  =0, & \left[  C_{i},C_{\left[
1,3\right]  }\right]  =0, & \left[  C_{i},C_{\left[  2,2\right]  }\right]
=0,\\
\left[  C_{\left[  1,2\right]  },C_{\left[  1,3\right]  }\right]  =0, &
\left[  C_{\left[  1,2\right]  },C_{\left[  2,2\right]  }\right]  =0, &
\left[  C_{\left[  2,2\right]  },C_{\left[  1,3\right]  }\right]  =0.
\end{array}
\end{equation}

In this case, the number of terms before symmetrization is $36,48$ and $72$
for $C_{\left[  1,2\right]  },C_{\left[  1,3\right]  }$ and $C_{\left[
2,2\right]  }$, respectively. Thus  the set $\left\{  H_{1},H_{2},H_{3}%
,C_{2},C_{3},C_{4},C_{\left[  2,1\right]  },C_{\left[  1,3\right]
},C_{\left[  2,2\right]  }\right\}  $ is complete formed by commuting
operators. The main objetctive of the KGR program is to find suitable linear
combinations of the three Racah operators to describe physical properties like
ionization energy or magnetic susceptibility, as well as to obtain information
on the stability island among the superheavy nuclei. This identification is
the second step after the simultaneous diagonalization of these operators, and
is heavily of numerical nature. The computation of the corresponding
eigenvalues for irreducible representations (IRREPs) of $\frak{su}(2,2)$
constitutes the essential step to be compared with the existing experimental
data. This task is in progress.

\section{Final remarks}

It seems natural to think, whenever we are confronted with a (non-canonical) embedding
of Lie algebras and the corresponding MLP which is not multiplicity free, the
information lost is somehow determined by the chain itself, and
not by a priori external techniques. In this sense, the missing
label operators which arise from the contraction $\frak{g}$ should
correspond to a natural choice of physical labelling operators,
as they are obtained using only the available information on the
algebra-subalgebra chain, their invariants and the induced decomposition.
This suggests that these could be the correct physical operators to be considered for
the labelling of states. An argument supporting this
interpretation is the equivalence of the contraction procedure
with the K-matrix method in the Elliott model $\frak{su}(3)\supset\frak{so}(3)$
chain or the supermultiplet model. Whether the remaining
possibilities that arise from the general algebraic solution of
the missing label problem are physically more relevant than those
operators found by contraction, remains a question that should be
analyzed for any specific physical situation. In all  examples analyzed,
the contraction method provides at most $n$ of the $2n$
available operators, thus induces a kind of partition in the set
of labelling operators. This suggests the existence of a certain
kind of hierarchy among these operators, as well as the fact that
some of them are not directly related to the properties of the
embedding of the subalgebra, and therefore not equivalent to
these. The next natural step is to analyze if the contraction
$\frak{g}$ can also be used to derive the eigenvalues of the
missing label operators.

\medskip

We have shown that many missing label problems relevant to Physics
can be completely solved by using the properties of the reduction
chain $\frak{s}\supset\frak{s}^{\prime}$, by means of a Lie
algebra contraction associated to this reduction or the decomposition
induced on the Casimir operators of the original Lie algebra. Analyzing the
set of invariants of the involved Lie algebras, suitable
commuting operators can be found that solve the missing label
problem. From this perspective, the operators found inherit an
intrinsic meaning, namely as terms of the Casimir operators
of $\frak{s}$ being re-scaled by the contraction, up to some
combination of lower order invariants of $\frak{s}$ and
$\frak{s}^{\prime}$. We have recomputed some classical reductions
appearing in atomic and nuclear physics, obtaining complete
agreement with the result obtained by different authors and
techniques. We believe to we also have furnished a natural explanation of the
order of these operators, which are directly related to the order
of the Casimir operators of the contracting Lie algebra. For the
special case of $n=m=0$, a direct relation among
the invariants of $\frak{s}$ and $\frak{s}^{\prime}$ with those of
the contraction $\frak{g}$ has been observed.

\medskip

The generalized contraction method is useful to solve the MLP when the
number of invariants of the contraction associated to the
reduction chain $\frak{s}\supset \frak{s}^{\prime}$ exceeds the
number of needed commuting labelling operators. In the case where the
invariants do not suffice to find
a complete solution of the missing label problem, it is expectable
that labelling operators of the same degree appear. This suggests
that further terms of the Casimir operators of $\frak{s}$ that
disappear during the contraction can be useful to complete the set
of missing label operators. We have thus introduced a decomposition
of the Casimir operators, the terms of
which are all constitute solutions to the labelling problem.
From these terms a set of $n$
independent labelling operators can be extracted, reducing the
problem to determine which (linear) combinations are mutually orthogonal.
In this sense, the method proposed in \cite{C72} is
 a first approximation to solve the MLP using the properties
of reduction chains turns out to be useful in many
practical cases. The bi-degree of the Casimir operators of a Lie
algebra with respect to the variables associated to the generators
of a subalgebra are further a relevant tool to obtain and
classify these labelling operators, although further distinction
of terms, for example when the subalgebra consists of various
copies, is also convenient to deduce additional
operators.\footnote{This has been done for the chain
$\frak{su}(4)\supset \frak{su}(2)\times\frak{su}(2)$ \cite{El} for the Wigner supermultiplet and the
isospin-strange spin classification schemes, although this supplementary
subdivision cannot be deduced using only the bi-degree defined by the contraction.}
We remark that this additional labelling
depends on each particular, case, since it is related to the distinction of the generators of
the subalgebra and on their possible splitting into direct sums.

\medskip

Certain aspects related to the decomposition method of Casimir
operators based on the contractions and its use in labelling
problems are specially emphasized:

\begin{itemize}
\item  The solutions agree with the ``natural"  choice for the
labelling operators. Their interpretation as ``broken" Casimir
operators confers them a physical meaning, in contrast to some
operators obtained by other techniques, where the physical
content of the operator is sometimes not entirely clear.

\item  The decomposition provides also a consistent explanation to
the question why for a number of reduction chains the labelling
operators must have the same degree. This fact is related to the number of
terms when the decomposition is applied.

\item This could probably explain why the eigenvalues of such
labelling operators are not integers, as already indicated by
Racah \cite{Ra} and verified in some models. The decomposition implies that the
eigenvalues of the labelling operators contribute to the
eigenvalues of the Casimir operators, being parts of them. In this context, the
interpretation of a labelling operator as ``broken" Casimir
operator leads to the idea of ``broken" integer eigenvalues.
\end{itemize}

Some questions still remain open. For example, whether there exist
reductions $\frak{s}\supset \frak{s}^{\prime}$ for which the
method followed here provides all available labelling operators.
An answer in this direction implies to find the general solution
to the MLP for each considered chain. At the present time, only for a few
number of algebras these computations have been carried out
completely \cite{El,Pa}. A complete study of all physically
relevant reduction chains involving simple Lie algebras up to some
fixed rank will certainly provide new insights to this problem.
We have also observed the existence of  reduction
chains where the terms of
the decomposition are not sufficient to construct a set of
orthogonal labelling operators. This means that the requirement that
they commute is not directly related to the functional independence
of these operators. How to compute these operators and their relation
with the original embedding is still an open problem that must be analyzed.
Another similar problem, which is being analyzed by the author, is to obtain
complete sets of commuting Racah operators for all simple Lie algebras,
using the labelling problem determined by the Cartan subalgebra. Such
types of reduction have been shown to be of interest in algebraic models
of molecular physics and nuclear spectroscopy \cite{Oss}, and a systematized
approach would certainly be a step forward in their study.

\subsection*{Acknowledgments}
This work was partially supported by the research project
MTM2006-09152 of the Ministerio de Educaci\'on y Ciencia and
the project CCG07-UCM/ESP-2922 of the U.C.M. The
author expresses his gratitude to R. Ferraro and M. A. Castagnino
of the Instituto de Astronom\'{\i}a y F\'{\i}sica del Espacio of the
UBA, where part of this work was done.

\end{document}